\documentclass[aps,prmaterials,reprint,superscriptaddress,amsmath,amssymb]{revtex4-2}

\usepackage{graphicx}
\usepackage{dcolumn}
\usepackage{bm}
\usepackage{color}

\usepackage[utf8]{inputenc}
\usepackage[T1]{fontenc}
\usepackage{mathptmx}
\usepackage{etoolbox}
\usepackage[hidelinks]{hyperref}

\usepackage[version=3]{mhchem}

\DeclareUnicodeCharacter{0393}{\ensuremath{\Gamma}}
\DeclareUnicodeCharacter{03B1}{\ensuremath{\alpha}}
\DeclareUnicodeCharacter{03C0}{\ensuremath{\pi}}

\begin{document}

\author{Tammo van der Heide}
\author{B\'alint Aradi}
\affiliation{Bremen Center for Computational Materials Science, University of Bremen, Bremen, Germany}

\author{Ben Hourahine}
\affiliation{SUPA, Department of Physics, The University of Strathclyde, Glasgow, G4 0NG, United Kingdom}

\author{Thomas Frauenheim}
\affiliation{Constructor University, School of Science, Campus Ring 1, Bremen, Germany}
\affiliation{Computational Science and Applied Research Institute (CSAR), 518110, Shenzhen, China}
\affiliation{Beijing Computational Science Research Center (CSRC), 100193, Beijing, China}

\author{Thomas A.\ Niehaus}
\email{thomas.niehaus@univ-lyon1.fr}
\affiliation{Univ Lyon, Universit\'e Claude Bernard Lyon 1, CNRS, Institut Lumi\`ere Mati\`ere, F-69622, Villeurbanne, France}

\title{Hybrid functionals for periodic systems \\ in the density functional tight-binding method}

\date{\today}

\begin{abstract}
Screened range-separated hybrid (SRSH) functionals within generalized Kohn-Sham density functional theory (GKS-DFT) have been shown to restore a general $1/(r\varepsilon)$ asymptotic decay of the electrostatic interaction in dielectric environments.
Major achievements of SRSH include an improved description of optical properties of solids and correct prediction of polarization-induced fundamental gap renormalization in molecular crystals.
The density functional tight-binding method (DFTB) is an approximate DFT that bridges the gap between first principles methods and empirical electronic structure schemes.
While purely long-range corrected RSH are already accessible within DFTB for molecular systems, this work generalizes the theoretical foundation to also include screened range-separated hybrids, with conventional pure hybrid functionals as a special case.
The presented formulation and implementation is also valid for periodic boundary conditions (PBC) beyond the $\Gamma$-point.
To treat periodic Fock exchange and its integrable singularity in reciprocal space, we resort to techniques successfully employed by DFT, in particular a truncated Coulomb operator and the minimum image convention.
Starting from the first principles Hartree-Fock operator, we derive suitable expressions for the DFTB method, using standard integral approximations and their efficient implementation in the DFTB+ software package.
Convergence behavior is investigated and demonstrated for the polyacene series as well as two- and three-dimensional materials.
Benzene and pentacene molecular and crystalline systems show the correct polarization-induced gap renormalization by SRSH-DFTB at heavily
reduced computational cost compared to first principles methods.
\end{abstract}

\maketitle

\section{Introduction}\label{sec:introduction}
As a semi-empirical method, density functional tight binding (DFTB)~\cite{dftb1, scc-dftb} fills the gap between methods such as Hartree-Fock~\cite{hartreeHF, fockHF} or Kohn-Sham density functional theory (DFT)~\cite{dft, ks-dft} and fully empirical force-fields in the domain of computational chemistry, condensed matter physics and materials science.
Its high ratio of accuracy to computational cost renders DFTB well-suited for extended systems with large unit cells or long timescale molecular dynamics (MD).
Over the last two decades, the original DFTB formalism by \citeauthor{dftb1}~\cite{dftb1} has been expanded by a number of extensions, including self-consistent charge SCC-DFTB~\cite{scc-dftb} and its extension to third order DFTB3~\cite{dftb3}, spin and spin-orbit interactions~\cite{doi:10.1021/jp068802p}, time dependent TD-DFTB~\cite{td-dftb}, real-time rTD-DFTB using propagation of the reduced one body density matrix and Ehrenfest dynamics~\cite{rtehrenfest1, rtehrenfest2}, machine learning enhanced repulsive potentials~\cite{chimes, fortnet} as well as non-equilibrium Green's function based electron transport~\cite{green1}.

For molecular calculations, range-separated hybrid functionals (RSH)~\cite{rsh, rsh1, rsh2, cam, rsh4} which constitute a mixture of non-local Fock like and (semi-)local exchange of DFT have been established as standard technique to address the inherent electronic self-interaction error (SIE)~\cite{sie} of DFT and restore the piecewise-linear~\cite{piecewise-linear, piecewise-linear2} behavior of the exact exchange-correlation functional between integer occupations.
Niehaus and Della Sala~\cite{rsh-dftb-formalism} generalized the DFTB formalism to non-periodic long-range corrected hybrid functionals (LC-DFTB), based on GKS-DFT and the density matrix as basic  variable in the expansion of the Kohn-Sham energy functional, which was later implemented in the DFTB+~\cite{rsh-dftb-implementation, dftb+} software package.

For periodically repeating structures, difficulties due to the Coulomb singularity of Fock exchange initially prevented an immediate and widespread adoption of hybrid functionals for solids.
The CRYSTAL~\cite{crystal} software package, based on the work of \citeauthor{early-fexchange}~\cite{early-fexchange}, provided the first publicly available implementation of periodic Fock exchange, paving the way for making periodic hybrid functionals readily accessible for solids.
In recent years effort has been made to develop reliable schemes for treating the singularity, including pioneering work by Gygi and Baldereschi~\cite{singularityAuxfunc} who lifted the singularity by introducing auxiliary functions, Spencer and Alavi~\cite{hfx-ct} resorting to a truncated Coulomb operator that is relatively simple to implement and does not posses a singularity in reciprocal space, and most recently Sundararaman and Arias'~\cite{ws-truncation} analytical proof of Wigner-Seitz truncation as an ideal method for regularizing the Coulomb potential in the exchange kernel.

General range-separated hybrid functional implementations have been developed for plane-wave~\cite{hfx-LAPW-1, hfx-LAPW-2, hfx-PAW-1, hfx-PAW-2, hfx-pw-1} and localized numerical~\cite{hfx-all-electron-numeric, fhiaims-1, fhiaims-2} or Gaussian-type~\cite{hfx-gto, hfx-gto-1, algo-ct, hfx-mic} orbital based DFT codes. Here we derive suitable expressions for the DFTB method and implement a real-space formulation of periodic Fock exchange in the DFTB+ software package.
We compare two algorithmic solutions that follow the footsteps of the truncated Coulomb interaction (TCI) by \citeauthor{algo-ct}~\cite{algo-ct} as well as the minimum image convention (MIC) by \citeauthor{hfx-mic}~\cite{hfx-mic} that utilizes the full, unscreened, Coulomb interaction.

The present work is structured as follows.
In Section~\ref{sec:theory}, we outline the basic periodic hybrid-functional DFTB theory, starting from the expansion of the total energy functional, deriving the periodic Fock exchange Hamiltonian within DFTB (by imposing Born--von K\'arm\'an (BvK) periodic boundary conditions for the density matrix of a finite mesh of k-points).
Section~\ref{sec:implementation} translates the equations into a formulation that is compatible with the neighbor list concept of DFTB+, and additionally gives expressions optimized for $\mathrm{\Gamma}$-point only k-point sampling, enabling efficient simulation of large systems with thousands of atoms.
This Section also contains benchmarks covering the scaling of the $\mathrm{\Gamma}$-point implementation with system size and the parallel performance of the k-point implementation.
Total energy and band-gap convergence behavior is then investigated and demonstrated by Section~\ref{sec:convergence} for the polyacene series, complemented by armchair graphene nanoribbons, two-dimensional h-BN monolayers and GaAs bulk in the Supplemental Material~\cite{supplmat}.
In Section~\ref{sec:gaprenorm}, benzene and pentacene molecular and crystalline systems are shown to exhibit the correct polarization-induced gap renormalization, that occurs if the surrounding dielectric medium is properly taken into account.
We close this work with a summary of our findings and by providing a brief outlook in Section~\ref{sec:s&o}.

We should note that the focus of the present work lies in the theory and implementation of the method, rather than an in-depth benchmarking of its accuracy.
The latter will be the topic of a forthcoming article.

\section{Theory}\label{sec:theory}
\subsection{Periodic GKS-DFT formalism}\label{sec:ks-dft-expansion}
In the following we outline the basic periodic KS-DFT formalism, including Fock exchange. This provides the notation and forms the basis for the DFTB approach described in Section~\ref{sec:dftb}.
All quantities are given in atomic units throughout and we will denote the crystal momentum as $\bm{k}$ and $\bm{k}'$, while real space lattice vectors are denoted as $\bm{g}$, $\bm{h}$, $\bm{l}$ and $\bm{m}$.

According to the Coulomb-attenuating method (CAM)~\cite{cam}, based on pioneering works by Gill~\cite{gill} and Savin~\cite{savin}, the electron-electron interaction is partitioned into short- and long-range contributions using the adiabatic connection theorem~\cite{adiabatictheorem}
\begin{align}\label{eq:cam}
\frac{1}{r} = \underbrace{\frac{1 - (\alpha + \beta) + \beta \mathrm{e}^{-\omega r}}{r}}_{\text{DFT}} + \underbrace{\frac{\alpha + \beta (1 - \mathrm{e}^{-\omega r})}{r}}_{\text{HF}},
\end{align}
where the short-range part is handled by a modified purely density functional, that ensures mutual error cancellation of local exchange and correlation, while the exact Fock exchange enforces the correct asymptotic decay in the long-range limit.
The parameters $\alpha, \beta$ and $\omega$ determine the fraction of global and long-range exact Fock exchange, as well as the value of the smooth range-separation function, which we have assumed to be of Yukawa type.

Complementing the usual kinetic energy $T_0[\rho]$ of the auxiliary
system of non-interacting electrons in GKS-DFT with the classical
Coulomb interaction $E_\mathrm{H}[\rho]$ (Hartree term), external
potential $v^\mathrm{ext}(\bm{r})$ and nuclear-repulsion energy
$E_\mathrm{NN}$, the above partitioning leads to the total energy
expression \emph{per unit cell} (UC)
\begin{align}\label{eq:ks-energy-functional}
E[\rho] &= T^0[\rho] + \frac{1}{N} \int v^\mathrm{ext}(\bm{r})\rho(\bm{r})\,\mathrm{d}\bm{r} + E_\mathrm{H}[\rho] + E_\mathrm{NN} + E_\mathrm{c}^\mathrm{DFT} \notag
\\
&+ [1 - (\alpha + \beta)]E_\mathrm{x}^\mathrm{DFT} + \alpha E_\mathrm{x}^\mathrm{HF} + \beta (E_{\mathrm{x},\mathrm{sr}}^{\omega,\mathrm{DFT}} + E_{\mathrm{x},\mathrm{lr}}^{\omega,\mathrm{HF}}),
\end{align}
with $N$ the total number of such cells in the crystal. This expression can be shortened by introducing the local part of the xc-functional
\begin{align}
E_\mathrm{xc,loc}[\rho] &:= [1 - (\alpha + \beta)]E_\mathrm{x}^\mathrm{DFT} + \beta E_{\mathrm{x},\mathrm{sr}}^{\omega,\mathrm{DFT}} + E_\mathrm{c}^\mathrm{DFT}.
\end{align}
Note that $E_\mathrm{xc,loc}[\rho] = E_\mathrm{xc,loc}^{\alpha ,\beta ,\omega}[\rho]$, but we refrain from explicitly stating the additional parameter dependencies for brevity.
Expressing the individual contributions of
Eq.\,\eqref{eq:ks-energy-functional} in terms of orbitals
$\psi_{i\bm{k}}(\bm{r})$, with additional quantum number $\bm{k}$, the
total energy reads
\begin{align}\label{eq:ks-energy-explicit}
E &=  \sum_{\bm{k}} w_{\bm{k}} \sum_i f_{i\bm{k}}
    \int\psi^*_{i\bm{k}}(\bm{r}) \left[- \frac{\Delta}{2} + v^\mathrm{ext}(\bm{r})  \right] \psi_{i\bm{k}}(\bm{r}) \,\mathrm{d}\bm{r} \notag
\\
&+ \frac{N}{2} \sum_{\bm{k},\bm{k}'} w_{\bm{k}}
     w_{\bm{k}'}\sum_{ij} f_{i\bm{k}} f_{j\bm{k}'} \iint \frac{|\psi_{i\bm{k}}(\bm{r})|^2|\psi_{j\bm{k}'}(\bm{r}')|^2}{|\bm{r} - \bm{r}'|}\,\mathrm{d}\bm{r}\mathrm{d}\bm{r}' \notag
\\
&- \frac{N}{4} \sum_{\bm{k},\bm{k}'} w_{\bm{k}}  w_{\bm{k}'} \sum_{ij} f_{i\bm{k}} f_{j\bm{k}'} \iint \mathrm{d}\bm{r} \mathrm{d}\bm{r}' \notag
\\
&\times \psi^*_{i\bm{k}}(\bm{r}) \psi_{j\bm{k}'}(\bm{r})\frac{ \alpha + \beta \left[ 1 - \mathrm{e}^{-\omega|\bm{r} - \bm{r}'|} \right] }{|\bm{r} - \bm{r}'|} \psi^*_{j\bm{k}'}(\bm{r}') \psi_{i\bm{k}}(\bm{r}') \notag
\\
&+ E_\mathrm{xc,loc}[\rho] + E_\mathrm{NN}.
\end{align}  
We describe a spin-unpolarized formalism for closed shell systems, where the spin degrees of freedom have already been summed up.
The occupation of eigenstate $i$ at crystal momentum $\bm{k}$ is denoted as $f_{i\bm{k}} \in [0, 2]$.
The number of states $i$ is given by the number of basis functions in the UC.
We further introduce weights $w_{\bm{k}}$, with normalization $\sum_{\bm{k}} w_{\bm{k}} = 1$, that arise from sampling the Brillouin zone by only selecting a subset of all wave-vectors compatible with the BvK cell.

For a crystal that is invariant with respect to cell translations, the GKS-orbitals are expected to obey the Bloch theorem and extend throughout the whole crystal.
We introduce Bloch-functions $\beta_\mu^{\bm{k}}(\bm{r})$, that emerge from a unitary transformation of the atomic orbitals $\phi_{\mu}(\bm{r})$ for orbital $\mu$ centered on an atom in the reference cell, when shifted by any real-space lattice vector $\bm{g}$:
\begin{align}\label{eq:bloch-functions}
\beta_\mu^{\bm{k}}(\bm{r}) = \frac{1}{\sqrt{N}} \sum_{\bm{g}} \phi_{\mu}(\bm{r} - \bm{g}) \mathrm{e}^{i\bm{k} \cdot \bm{g}}.
\end{align}

Expressing the wavefunctions as a linear combination of crystalline orbitals leads to states that equally possess Bloch-wave character, therefore satisfying the Bloch-condition $\psi_{i\bm{k}}(\bm{r} + \bm{g}) = \psi_{i\bm{k}}(\bm{r}) \mathrm{e}^{i\bm{k} \cdot \bm{g}}$
\begin{align}\label{eq:ks-orbitals}
\psi_{i\bm{k}}(\bm{r}) &= \sum_{\mu} c_{\mu i}(\bm{k}) \beta_\mu^{\bm{k}}(\bm{r}),
\end{align}
with eigenvector coefficients $c_{\mu i}(\bm{k})$ attributed to orbital $\mu$ and eigenstate $i$.
By exploiting the translational symmetry, $\hat{\mathrm{O}}(\bm{r} + \bm{g}, \bm{r}' + \bm{g}) = \hat{\mathrm{O}}(\bm{r}, \bm{r}')$, of an operator $\hat{\mathrm{O}}$, folding operations for transitions from direct to reciprocal space (and vice versa) are obtained:
\begin{align}
O_{\mu\nu}(\bm{g}) &= \langle \phi_\mu(\bm{r} - \bm{g}) | \hat{\mathrm{O}} | \phi_\nu(\bm{r}) \rangle = O_{\nu\mu}(-\bm{g}) \label{eq:Omn-rspace}
\\
O_{\mu\nu}(\bm{k}) &= \sum_{\bm{g}} O_{\mu\nu}(\bm{g}) \mathrm{e}^{-i \bm{k} \cdot \bm{g}}. \label{eq:Omn-to-kspace}
\end{align}
The special case $\hat{\mathrm{O}} = \mathbf{1}$ refers to the overlap matrix elements $S_{\mu\nu}$.
We introduce the convention that the real-space shifts in the arguments of Hamiltonian and overlap refer to the first orbital, while the second remains in the reference cell.
In reciprocal space, the density matrix is built from the eigenvector coefficients and occupations
\begin{align}
P_{\mu\nu}(\bm{k}) &= \sum_i f_{i\bm{k}} c_{\mu i}(\bm{k}) c_{\nu i}^*(\bm{k}) \label{eq:Pmn-kspace}
\\
P_{\mu\nu}(\bm{k}) &= \sum_{\bm{g}} P_{\mu\nu}(\bm{g}) \mathrm{e}^{-i \bm{k} \cdot \bm{g}} \label{eq:Pmn-to-kspace}
\\
P_{\mu\nu}(\bm{g}) &= \sum_{\bm{k}} w_{\bm{k}} P_{\mu\nu}(\bm{k}) \mathrm{e}^{i \bm{k} \cdot \bm{g}}, \label{eq:Pmn-to-rspace}
\end{align}
while transformations according to Eq.\,\eqref{eq:Omn-rspace} and Eq.\,\eqref{eq:Omn-to-kspace} apply.

For the two-electron, four-center integrals we resort to Mulliken's notation and distinguish between an unscreened and screened, long-range Coulomb kernel
\begin{align}\label{eq:4center-integrals}
&\left( \phi_\mu^{\bm{l}} \phi_\nu^{\bm{h}} | \phi_\kappa^{\bm{m}}
                                            \phi_\lambda^{\bm{g}}
                                            \right) :=
                                            \iint \mathrm{d}\bm{r}\mathrm{d}\bm{r}' \notag
\\
&\times \phi_\mu^*(\bm{r} - \bm{l}) \phi_\nu(\bm{r} - \bm{h}) \frac{1}{|\bm{r} - \bm{r}'|} \phi_\kappa^*(\bm{r}' - \bm{m}) \phi_\lambda(\bm{r}' - \bm{g})
\\
&\left( \phi_\mu^{\bm{l}} \phi_\nu^{\bm{h}} | \phi_\kappa^{\bm{m}} \phi_\lambda^{\bm{g}} \right)^\mathrm{lr,\omega} := \iint \mathrm{d}\bm{r}\mathrm{d}\bm{r}' \notag
\\
&\times \phi_\mu^*(\bm{r} - \bm{l}) \phi_\nu(\bm{r} - \bm{h}) \frac{1 - \mathrm{e}^{-\omega|\bm{r} - \bm{r}'|}}{|\bm{r} - \bm{r}'|} \phi_\kappa^*(\bm{r}' - \bm{m}) \phi_\lambda(\bm{r}' - \bm{g}),
\end{align}
with shorthand notation $\phi_\mu^{\bm{g}} := \phi_\mu(\bm{r} - \bm{g})$.
We represent the total energy of Eq.\,\eqref{eq:ks-energy-explicit} in terms of the new Bloch basis and exploit the translational symmetry of an arbitrary (in general, non-local) operator in direct space to choose the reference (zeroth) cell and carry out a trivial lattice summation, resulting in
\begin{align}\label{eq:ks-energy-bloch}
E &= \sum_{\bm{k}} w_{\bm{k}} \sum_{\mu\nu} h_{\mu\nu}(\bm{k}) P_{\nu\mu}(\bm{k}) + E_\mathrm{xc,loc}[\rho] + E_\mathrm{NN} \notag
\\
&+ \frac{1}{2} \sum_{\bm{k}\bm{k}'} w_{\bm{k}} w_{\bm{k}'} \sum_{\mu\nu\kappa\lambda} P_{\nu\mu}(\bm{k}) P_{\lambda\kappa}(\bm{k}') \notag
\\
&\times \sum_{\bm{g}\bm{h}\bm{m}} \mathrm{e}^{i \bm{k}' (\bm{g} - \bm{m})} \left( \phi_\mu^{\bm{0}} \phi_\nu^{\bm{h}} | \phi_\kappa^{\bm{m}} \phi_\lambda^{\bm{g}} \right) \mathrm{e}^{i \bm{k} \cdot \bm{h}} \notag
\\
&- \frac{1}{4} \sum_{\bm{k}\bm{k}'} w_{\bm{k}} w_{\bm{k}'} \sum_{\mu\nu\kappa\lambda} P_{\nu\mu}(\bm{k}) P_{\lambda\kappa}(\bm{k}') \sum_{\bm{g}\bm{h}\bm{m}} \mathrm{e}^{-i \bm{k}' (\bm{m} - \bm{h})} \notag
\\
&\times \left[ \alpha \left( \phi_\mu^{\bm{0}} \phi_\lambda^{\bm{h}} | \phi_\kappa^{\bm{m}} \phi_\nu^{\bm{g}} \right) + \beta \left( \phi_\mu^{\bm{0}} \phi_\lambda^{\bm{h}} | \phi_\kappa^{\bm{m}} \phi_\nu^{\bm{g}} \right)^\mathrm{lr,\omega} \right] \mathrm{e}^{i \bm{k} \cdot \bm{g}}.
\end{align}
Here, $h_{\mu\nu}(\bm{k})$ refers to the one-electron integral matrix elements of the kinetic energy and external potential.

By applying the variational principle to Eq.\,\eqref{eq:ks-energy-bloch} we obtain the secular equation
\begin{align}\label{eq:dft-generalized-eigenproblem}
\sum_\nu c_{\nu i}(\bm{k})\left[ H_{\mu\nu}(\bm{k}) - \varepsilon_i S_{\mu\nu}(\bm{k}) \right] = 0,\quad \forall i
\end{align}
with $\bm{k}$-dependent Hamiltonian matrix elements
\begin{align}\label{eq:Hmn-generalized-eigenproblem}
H_{\mu\nu}(\bm{k}) &= h_{\mu\nu}(\bm{k}) + \sum_{\bm{g}} v_{\mu\nu}^\mathrm{xc,loc}(\bm{g}) \mathrm{e}^{-i \bm{k} \cdot \bm{g}} \notag
\\
&+ \sum_{\bm{k}'} w_{\bm{k}'} \sum_{\lambda\kappa} P_{\lambda\kappa}(\bm{k}') \sum_{\bm{g}\bm{h}\bm{m}} \mathrm{e}^{i \bm{k}' (\bm{g} - \bm{m})} \notag
\\
&\times \left( \phi_\mu^{\bm{0}} \phi_\nu^{\bm{h}} | \phi_\kappa^{\bm{m}} \phi_\lambda^{\bm{g}} \right) \mathrm{e}^{i \bm{k} \cdot \bm{h}} \notag
\\
&- \frac{1}{2} \sum_{\bm{k}'} w_{\bm{k}'} \sum_{\lambda\kappa} P_{\lambda\kappa}(\bm{k}') \sum_{\bm{g}\bm{h}\bm{m}} \mathrm{e}^{i \bm{k} \cdot \bm{g}} \mathrm{e}^{i \bm{k}' (\bm{h} - \bm{m})} \notag
\\
&\times \left[ \alpha \left( \phi_\mu^{\bm{0}} \phi_\lambda^{\bm{h}} | \phi_\kappa^{\bm{m}} \phi_\nu^{\bm{g}} \right) + \beta \left( \phi_\mu^{\bm{0}} \phi_\lambda^{\bm{h}} | \phi_\kappa^{\bm{m}} \phi_\nu^{\bm{g}} \right)^\mathrm{lr,\omega} \right],
\end{align}
where the local part of the exchange-correlation potential (i.e., the functional derivative of $E_{\mathrm{xc,loc}}$) is introduced as
\begin{equation}
  \label{vxc}
  v_{\mu\nu}^\mathrm{xc,loc}[\rho](\bm{g}) = \int \phi_\mu^{\bm{g}}(\bm{r})^* v_{\mathrm{xc,loc}}[\rho](\bm{r}) \phi_\nu^{\bm{0}}(\bm{r})\,\mathrm{d}\bm{r}.
\end{equation}

Bloch-functions are bases for the irreducible representations of the translation group and super-matrices in reciprocal space, such as $H_{\mu\nu}(\bm{k})$, can be transformed into a block-diagonal form by unitary transformation~\cite{crystal-vib}.
This gives rise to an independent diagonalization of Eq.\,\eqref{eq:dft-generalized-eigenproblem} for each different k-point.

\subsection{Density functional tight binding}\label{sec:dftb}
In the following we outline the periodic RSH-DFTB formalism of second order in the energy expansion.

The minimal valence-only basis set $\{\phi_{\mu}\}$ of atomic orbitals entering the eigenfunction \emph{ansatz} of Eq.\,\eqref{eq:ks-orbitals} is obtained by performing first principles RSH calculations for neutral and spin-unpolarized pseudo-atoms, as described in Refs.~\onlinecite{dftb3, rsh-dftb-implementation}.
We continue by approximating the local part of the exchange-correlation functional, by expanding around the reference up to second order in the perturbation
\begin{align}
&E_\mathrm{xc,loc}[\rho_0 + \delta\rho] = E_\mathrm{xc,loc}[\rho_0] +
                \frac{1}{N} \int v_\mathrm{xc,loc}[\rho_0](\bm{r}) \delta\rho(\bm{r})\mathrm{d}\bm{r} \notag
\\
&+ \frac{1}{2N} \iint f_\mathrm{xc,loc}[\rho_0](\bm{r},\bm{r}') \delta\rho(\bm{r})\delta\rho(\bm{r}')\,\mathrm{d}\bm{r}\mathrm{d}\bm{r}' + \mathcal{O} (\delta\rho^3) \label{eq:Exc-linearized},
\end{align}
thus linearizing the exchange-correlation potential.
The density matrix of Eq.\,\eqref{eq:Pmn-kspace} is decomposed into reference and perturbation, such that ${\bf P} = {\bf P}^{(0)} + \Delta {\bf P}$.
Usually, the reference ${\bf P}^{(0)} = \sum_A {\bf P}_A$ is constructed as a superposition of densities of non-interacting atoms, where the sum runs over all atoms ($A$) in the unit cell.
Representing Eq.\,\eqref{eq:Exc-linearized} in the Bloch basis of Eq.\,\eqref{eq:ks-orbitals} then yields
\begin{align}
E_\mathrm{xc,loc}[\rho] &\approx E_\mathrm{xc,loc}[\rho_0] + \sum_{\bm{k}} w_{\bm{k}} \sum_{\mu\nu}\Delta P_{\nu\mu}(\bm{k}) v_{\mu\nu}^\mathrm{xc,loc}(\bm{k}) \notag
\\
&+ \frac{1}{2} \sum_{\bm{k}\bm{k}'} w_{\bm{k}} w_{\bm{k}'} \sum_{\mu\nu\kappa\lambda} \Delta P_{\nu\mu}(\bm{k}) \Delta P_{\lambda\kappa}(\bm{k}') \notag
\\
&\times \sum_{\bm{g}'} \sum_{\bm{l}\bm{l}'} \mathrm{e}^{i \bm{k} \cdot \bm{g}'} \mathrm{e}^{i \bm{k}' (\bm{l}' - \bm{l})} f_{\mu\nu\kappa\lambda}^\mathrm{xc,loc}(\bm{0},\bm{g}',\bm{l},\bm{l}') \label{eq:Exc-linearized-bloch}
\\
f_{\mu\nu\kappa\lambda}^\mathrm{xc,loc}(\bm{g},\bm{g}',\bm{l},\bm{l}')
                        &:= \iint \mathrm{d}\bm{r}\mathrm{d}\bm{r}' \notag
\\
&\times \phi_\mu^{\bm{g}}(\bm{r})^* \phi_\nu^{\bm{g}'}(\bm{r}) f_{\mathrm{xc,loc}}(\bm{r},\bm{r}') \phi_\kappa^{\bm{l}}(\bm{r}')^* \phi_\lambda^{\bm{l}'}(\bm{r}'),
\end{align}
where the matrix elements of first and second functional derivatives are denoted as $v_{\mu\nu}^\mathrm{xc,loc}$ and $f_{\mu\nu\kappa\lambda}^\mathrm{xc,loc}$, respectively.

We continue by inserting Eq.\,\eqref{eq:Exc-linearized-bloch} into the original energy functional of Eq.\,\eqref{eq:ks-energy-bloch}, leading to
\begin{align}\label{eq:E-dftb}
E &= \sum_{\bm{k}} w_{\bm{k}} \sum_{\mu\nu} H^{(0)}_{\mu\nu}(\bm{k})
    P_{\nu\mu}(\bm{k}) + E^{(2)}_\text{DFT} +
    E^{(2)}_\text{HF} + E_\mathrm{rep},
\end{align}
with $E_\mathrm{rep}$ covering all terms that solely depend on the reference density and $E^{(2)}_\text{DFT}$ and $E^{(2)}_\text{HF}$ being of second order in $\Delta {\bf P}$
\begin{align}
E^{(2)}_\text{DFT} &=
\frac{1}{2} \sum_{\bm{k}\bm{k}'} w_{\bm{k}} w_{\bm{k}'} \sum_{\mu\nu\lambda\kappa} \Delta P_{\nu\mu}(\bm{k}) \Delta P_{\lambda\kappa}(\bm{k}') \sum_{\bm{g}\bm{h}\bm{m}} \mathrm{e}^{i \bm{k} \cdot \bm{h}} \notag
\\
&\times \mathrm{e}^{i \bm{k}' (\bm{g} - \bm{m})} \left[ \left( \phi_\mu^{\bm{0}} \phi_\nu^{\bm{h}} | \phi_\kappa^{\bm{m}} \phi_\lambda^{\bm{g}} \right) + f_{\mu\nu\kappa\lambda}^\mathrm{xc,loc}(\bm{0},\bm{h},\bm{m},\bm{g}) \right] \label{eq:E2-bloch-dft}
\\
E^{(2)}_\text{HF} &= \alpha E^{x,\mathrm{fr}} + \beta E^{x,\mathrm{lr}}
\\
&= - \frac{1}{4} \sum_{\bm{k}\bm{k}'} w_{\bm{k}} w_{\bm{k}'} \sum_{\mu\nu\lambda\kappa} \Delta P_{\nu\mu}(\bm{k}) \Delta P_{\lambda\kappa}(\bm{k}') \sum_{\bm{g}\bm{h}\bm{m}} \mathrm{e}^{i \bm{k} \cdot \bm{g}} \notag
\\
&\times \left[ \alpha \left( \phi_\mu^{\bm{0}} \phi_\lambda^{\bm{h}} | \phi_\kappa^{\bm{m}} \phi_\nu^{\bm{g}} \right) + \beta \left( \phi_\mu^{\bm{0}} \phi_\lambda^{\bm{h}} | \phi_\kappa^{\bm{m}} \phi_\nu^{\bm{g}} \right)^\mathrm{lr,\omega} \right] \mathrm{e}^{i \bm{k}' (\bm{h} - \bm{m})}. \label{eq:E2-bloch-hf}
\end{align}

Further, Eq.\,\eqref{eq:E-dftb} introduced the zeroth-order Hamiltonian $H^{(0)}_{\mu\nu}(\bm{k})$, that is defined as follows
\begin{align}\label{eq:Hmn0}
H_{\mu\nu}^{(0)}(\bm{k}) &:= h_{\mu\nu}(\bm{k}) + \sum_{\bm{g}}
                           v_{\mu\nu}^{\mathrm{xc,loc}}[\rho_0](\bm{g}) \mathrm{e}^{i \bm{k} \cdot \bm{g}} \notag
\\
&+ \sum_{\bm{k}'} w_{\bm{k}'} \sum_{\lambda\kappa} P^{(0)}_{\lambda\kappa}(\bm{k}') \sum_{\bm{g}\bm{h}\bm{m}} \mathrm{e}^{i \bm{k}' (\bm{g} - \bm{m})} \notag
\\
&\times \left( \phi_\mu^{\bm{0}} \phi_\nu^{\bm{h}} | \phi_\kappa^{\bm{m}} \phi_\lambda^{\bm{g}} \right) \mathrm{e}^{i \bm{k} \cdot \bm{h}} \notag
\\
&- \frac{1}{2} \sum_{\bm{k}'} w_{\bm{k}'} \sum_{\lambda\kappa} P^{(0)}_{\lambda\kappa}(\bm{k}') \sum_{\bm{g}\bm{h}\bm{m}} \mathrm{e}^{i \bm{k} \cdot \bm{g}} \mathrm{e}^{i \bm{k}' (\bm{h} - \bm{m})} \notag
\\
&\times \left[ \alpha \left( \phi_\mu^{\bm{0}} \phi_\lambda^{\bm{h}} | \phi_\kappa^{\bm{m}} \phi_\nu^{\bm{g}} \right) + \beta \left( \phi_\mu^{\bm{0}} \phi_\lambda^{\bm{h}} | \phi_\kappa^{\bm{m}} \phi_\nu^{\bm{g}} \right)^\mathrm{lr,\omega} \right].
\end{align}
As usual in the DFTB framework, we adopt the two-center approximation and replace onsite-blocks in the Hamiltonian by diagonal matrices with free atom eigen-energies, $\epsilon_{\mu}^{\text{free}}$, ensuring the correct limit on dissociation and leading to
\begin{align}
\label{eq:h0}
  H^{(0)}_{\mu\nu}(\bm{g}) =
  \begin{cases}
    \epsilon^{\text{free}}_{\mu} & \mu = \nu \\
    H^{(0)}_{\mu\nu}[\rho_A + \rho_B] & \mu\in A,\ \nu\in B \\
    0 & \text{else}.
  \end{cases}
\end{align}
$\rho_A = \rho_A(\bm{r} - \bm{g})$ and $\rho_B = \rho_B(\bm{r})$ denote atomic densities, derived from appropriate pseudo-atom calculations.
$\epsilon_{\mu}^{\text{free}}$, as well as the non-diagonal elements of Eq.\,\eqref{eq:h0}, are obtained from RSH-DFT calculations and latter stored for high-symmetry orbital configurations as a function of distance between atoms $A$ and $B$ in Slater-Koster tables~\cite{sk-trans}.
The matrix elements $H_{\mu\nu}^{(0)}(\bm{k})$ are recovered by transforming according to Eq.\,\eqref{eq:Omn-to-kspace}.

In line with conventional DFTB and as generalized to periodic boundary conditions, the repulsive energy is approximated by a sum of fast decaying pair potentials~\cite{Erep-approx}
\begin{align}\label{eq:Erep-periodic}
E_\mathrm{rep} &= \frac{1}{2} \sum_{A,B}^\mathrm{UC} \sum_{\bm{g}} V_\mathrm{rep}^{AB}(\bm{R}_{AB} - \bm{g}),
\end{align}
either determined by a higher level of theory~\cite{dftb3} or fitted to empirical data~\cite{Erep-exp-fitted}.
The sums over atoms $A,B$ are restricted to the reference unit cell, while an additional direct lattice sum $\bm{g}$ also accounts for contributions of images in neighboring cells.

The term $E^{(2)}_\text{DFT}$, defined in Eq.\,\eqref{eq:E2-bloch-dft} is treated by applying the Mulliken approximation
\begin{align}
\phi_\mu^*(\bm{r})\phi_\nu(\bm{r}) \approx \frac{1}{2} S_{\mu\nu} \left( |\phi_\mu(\bm{r})|^2 + |\phi_\nu(\bm{r})|^2 \right)
\end{align}
to the four-center integrals
\begin{align}\label{eq:mulliken-approximation}
\left( \phi_\mu^{\bm{l}} \phi_\nu^{\bm{h}} | \phi_\lambda^{\bm{m}} \phi_\kappa^{\bm{g}} \right) &\approx \frac{1}{4} S_{\mu\nu}(\bm{l} - \bm{h}) S_{\lambda\kappa}(\bm{m} - \bm{g}) \notag
\\
& \times \Big[ \left( \phi_\mu^{\bm{l}} \phi_\mu^{\bm{l}} | \phi_\lambda^{\bm{m}} \phi_\lambda^{\bm{m}} \right) + \left( \phi_\mu^{\bm{l}} \phi_\mu^{\bm{l}} | \phi_\kappa^{\bm{g}} \phi_\kappa^{\bm{g}} \right) \notag
\\
&+ \left( \phi_\nu^{\bm{h}} \phi_\nu^{\bm{h}} | \phi_\lambda^{\bm{m}} \phi_\lambda^{\bm{m}} \right) + \left( \phi_\nu^{\bm{h}} \phi_\nu^{\bm{h}} | \phi_\kappa^{\bm{g}} \phi_\kappa^{\bm{g}} \right) \Big]
\\
&= \frac{1}{4} S_{\mu\nu}(\bm{l} - \bm{h}) S_{\lambda\kappa}(\bm{m} - \bm{g}) \notag
\\
&\times \Big[ \gamma_{\mu\lambda}(\bm{l} - \bm{m}) + \gamma_{\mu\kappa}(\bm{l} - \bm{g}) \notag
\\
&+ \gamma_{\nu\lambda}(\bm{h} - \bm{m}) + \gamma_{\nu\kappa}(\bm{h} - \bm{g}) \Big].
\end{align}
In line with the convention introduced by Eq.\,\eqref{eq:Omn-rspace}, the real-space shifts in the arguments of $\gamma$ are associated with the first orbital, while the second orbital remains in the central cell.
We follow the reasoning of conventional second-order DFTB, by approximating the orbital products $|\phi_\mu|^2$ as exponentially decaying spherically symmetric charge densities, leading to three integral parameterizations
\begin{align}
\gamma^\mathrm{fr}_{\mu\nu} &= \gamma_{AB}^\mathrm{fr}(R_{AB}) =
                              \frac{\tau^3_A\tau^3_B}{(8\pi)^2} \iint \mathrm{d}\bm{r} \mathrm{d}\bm{r}' \notag
\\
&\times \mathrm{e}^{-\tau_A |\bm{r} - \bm{R}_A|} \left[\frac{1}{|\bm{r} - \bm{r}'|} + f_\mathrm{xc,loc}[\rho_0]\right] \mathrm{e}^{-\tau_B |\bm{r}' - \bm{R}_B|}
\\
\gamma^\mathrm{fr,HF}_{\mu\nu} &= \gamma_{AB}^\mathrm{fr,HF}(R_{AB}) = \frac{\tau^3_A\tau^3_B}{(8\pi)^2} \iint \mathrm{d}\bm{r} \mathrm{d}\bm{r}' \notag
\\
&\times \mathrm{e}^{-\tau_A |\bm{r} - \bm{R}_A|} \frac{1}{|\bm{r} - \bm{r}'|} \mathrm{e}^{-\tau_B |\bm{r}' - \bm{R}_B|}
\\
\gamma^\mathrm{lr,HF}_{\mu\nu} &= \gamma_{AB}^\mathrm{lr,HF}(R_{AB}) = \frac{\tau^3_A\tau^3_B}{(8\pi)^2} \iint \mathrm{d}\bm{r} \mathrm{d}\bm{r}' \notag
\\
&\times \mathrm{e}^{-\tau_A |\bm{r} - \bm{R}_A|} \frac{1 - e^{-\omega|\bm{r} - \bm{r}'|}}{|\bm{r} - \bm{r}'|} \mathrm{e}^{-\tau_B |\bm{r}' - \bm{R}_B|},
\end{align}
with analytical expressions~\cite{scc-dftb, rsh-dftb-implementation}.
A distinction is made between screened long-range (lr) and unscreened full-range (fr) kernels.
The parameter $\tau_A$ is obtained from requiring RSH-DFT and RSH-DFTB yield the same second derivative of the total energy for an atom with respect to orbital occupation,
i.e., predicting the same chemical hardness~\cite{rsh-dftb-implementation}.
A detailed derivation is provided in Appendix~\ref{sec:hubbu-cam-dftb}.

Now we have all the prerequisites to treat the semi-local energy contribution $E^{(2)}_\text{DFT}$.
To this end we compute Mulliken populations, $q_\mu$, of orbital $\mu$
\begin{align}
q_\mu
&= \sum_{\bm{k}} w_{\bm{k}} \sum_{\nu} P_{\mu\nu}(\bm{k}) S_{\nu\mu}(\bm{k}) \\
&= \sum_{\bm{g}} \sum_{\nu} P_{\mu\nu}(\bm{g}) S_{\nu\mu}(\bm{g}), \label{eq:mulliken-rspace}
\end{align}
evaluated in direct or reciprocal space.
The net charges of atom $A$ are obtained by comparing the populations with the neutral atom ($Z_A$)
\begin{align}
\Delta q_A = \sum_{\mu[A]} \Delta q_\mu = q_A - Z_A.
\end{align}
The final result is obtained by summing over all atoms $A,B$ in the unit cell and accounting for any periodic images of $B$, as captured by the sum over direct lattice vectors $\bm{g}$ extending throughout the crystal
\begin{align}\label{eq:coulomb-energy-periodic}
E^{(2)}_\text{DFT} &= \frac{1}{2} \sum_{A,B}^\mathrm{UC} \sum_{\bm{g}} \gamma_{AB}^\mathrm{fr}(\bm{g}) \Delta q_A \Delta q_B.
\end{align}

\subsection{Periodic Fock exchange in DFTB}\label{sec:hfx}
To complete the theoretical foundation of periodic RSH-DFTB, the energy contributions due to the additional Fock terms require special treatment. We restrict the derivation to the screened, long-range Coulomb kernel, since its full-range counterpart is contained as the limiting case of a large value of the range-separation parameter, $\omega$.

Firstly, the (back) Fourier transformation of Eq.\,\eqref{eq:Pmn-to-rspace} is identified in order to simplify the expression, leading to
\begin{align}\label{eq:E-HFX-bloch-rspace}
E^{x,\mathrm{lr}} &= - \frac{1}{4} \sum_{\bm{k}} w_{\bm{k}} \sum_{\mu\nu\lambda\kappa} \sum_{\bm{g}\bm{h}\bm{l}} \Delta P_{\nu\mu}(\bm{k}) \Delta P_{\lambda\kappa}(-\bm{l}) \notag
\\
&\times \left(\phi_\mu^{\bf 0} \phi_\lambda^{\bm{h}} | \phi_\kappa^{\bm{h} + \bm{l}} \phi_\nu^{\bm{g}} \right)^\mathrm{lr,\omega} \mathrm{e}^{i\bm{k} \cdot \bm{g}}, \quad {\bm{l}} := \bm{m} - \bm{h}.
\end{align}
We now apply the Mulliken approximation and arrive at an expression that is compatible with the DFTB formalism
\begin{align}\label{eq:E-HFX-rspace-approx}
E^{x,\mathrm{lr}} &= -\frac{1}{16} \sum_{\bm{k}} w_{\bm{k}} \sum_{\mu\nu\lambda\kappa} \sum_{\bm{g}\bm{h}\bm{l}} \Delta P_{\nu\mu}(\bm{k}) \Delta P_{\lambda\kappa}(-\bm{l}) \notag
\\
& S_{\lambda\mu}(\bm{h}) S_{\kappa\nu}(\bm{l} + \bm{h} - \bm{g}) \Big[ \gamma_{\mu\nu}^\mathrm{lr,HF}(-\bm{g}) + \gamma_{\mu\kappa}^\mathrm{lr,HF}(-\bm{h} - \bm{l}) \notag
\\
&+ \gamma_{\lambda\nu}^\mathrm{lr,HF}(\bm{h} - \bm{g}) + \gamma_{\lambda\kappa}^\mathrm{lr,HF}(-\bm{l}) \Big] \mathrm{e}^{i\bm{k} \cdot \bm{g}}.
\end{align}
For later implementation using a neighbor list, as discussed in Section~\ref{sec:neigh-algo}, it is advantageous to carry out an index shift $\bm{l} \longrightarrow \bm{g} - \bm{h} - \bm{l}$ that simplifies the arguments of the real-space overlaps
\begin{align}\label{eq:E-HFX-rspace-approx-reindx}
E^{x,\mathrm{lr}} &= -\frac{1}{16} \sum_{\bm{k}} w_{\bm{k}} \sum_{\mu\nu\lambda\kappa} \sum_{\bm{g}\bm{h}\bm{l}} \Delta P_{\nu\mu}(\bm{k}) \Delta P_{\lambda\kappa}(\bm{h} - \bm{l} - \bm{g}) \notag
\\
& S_{\lambda\mu}(\bm{h}) S_{\kappa\nu}(\bm{l}) \Big[ \gamma_{\mu\nu}^\mathrm{lr,HF}(-\bm{g}) + \gamma_{\mu\kappa}^\mathrm{lr,HF}(-\bm{g} - \bm{l}) \notag
\\
&+ \gamma_{\lambda\nu}^\mathrm{lr,HF}(\bm{h} - \bm{g}) + \gamma_{\lambda\kappa}^\mathrm{lr,HF}(\bm{h} - \bm{l} - \bm{g}) \Big] \mathrm{e}^{i\bm{k} \cdot \bm{g}}.
\end{align}
By applying the variational principle and writing the above expression with respect to the density matrices, the corresponding ground-state Hamiltonian emerges as
\begin{align}\label{eq:dHmn-HFX-rspace-approx-reindx}
\Delta H^{x,\mathrm{lr}}_{\mu\nu}(\bm{k}) &= -\frac{1}{8} \sum_{\lambda\kappa} \sum_{\bm{g}\bm{h}\bm{l}} \Delta P_{\lambda\kappa}(\bm{h} - \bm{l} - \bm{g}) S_{\lambda\mu}(\bm{h}) S_{\kappa\nu}(\bm{l}) \notag
\\
&\times \Big[ \gamma_{\mu\nu}^\mathrm{lr,HF}(-\bm{g}) + \gamma_{\mu\kappa}^\mathrm{lr,HF}(-\bm{g} - \bm{l}) \notag
\\
&+ \gamma_{\lambda\nu}^\mathrm{lr,HF}(\bm{h} - \bm{g}) + \gamma_{\lambda\kappa}^\mathrm{lr,HF}(\bm{h} - \bm{l} - \bm{g}) \Big] \mathrm{e}^{i\bm{k} \cdot \bm{g}}.
\end{align}
This illustrates that the evaluation of the associated energy contribution becomes straightforward once the exchange Hamiltonian $\Delta H^{x,\mathrm{lr}}_{\mu\nu}(\bm{k})$ is available
\begin{align}\label{eq:E-rspace}
E^{x,\mathrm{lr}} &= \frac{1}{2} \sum_{\bm{k}} w_{\bm{k}} \sum_{\mu\nu} \Delta H^{x,\mathrm{lr}}_{\mu\nu}(\bm{k}) \Delta P_{\nu\mu}(\bm{k}).
\end{align}
Deriving the energy and Hamiltonian for global Fock-like exchange, instead of a range-separated expression, simply requires substituting $\gamma_{\mu\nu}^\mathrm{lr,HF}$ with $\gamma_{\mu\nu}^\mathrm{fr,HF}$.

\section{Implementation}\label{sec:implementation}
The extension of the DFTB formalism, according to Sections~\ref{sec:dftb} and \ref{sec:hfx}, requires modifications to a)
the parameterization suite \emph{skprogs}~\cite{skprogs-github} and b) the main DFTB+~\cite{dftb+} code. The newly developed routines are publicly available in a) the main branch of the official repository and b) a pull request of the respective development branch to the main branch of the official repository~\cite{periodicHybridsPR}, currently under code review by the maintainers.

For the zeroth-order Hamiltonian construction, we generalized the scheme of Lutsker and co-workers~\cite{rsh-dftb-implementation} to handle global Hartree-Fock exchange in addition to the already available screened kernels. This enables pre-tabulation of the $H^{(0)}$ and $S$ matrix elements for general CAM xc-functionals and further includes a generalized scheme to determine the decay constants according to Eq.\,\eqref{eq:U-cam-dftb}.

The central performance critical task of developing an efficient implementation to construct the Fock exchange matrix, as provided by Eq.\,\eqref{eq:dHmn-HFX-rspace-approx-reindx}, is based on the neighbor-list-based design of DFTB+. This utilizes the sparsity~\cite{dftb+old} pattern induced by the spatial decay of the real-space overlap matrix elements in the Slater-Koster tables.

\subsection{Neighbor list based algorithm}\label{sec:neigh-algo}
In order to re-formulate Eq.\,\eqref{eq:dHmn-HFX-rspace-approx-reindx}, utilizing the concept of neighbor lists, we first introduce the notation $\bar{L} \in \mathcal{N}(M)$ which refers to atom $\bar{L}$ being a neighbor of atom $M$, where atom $M$ is located in the central cell but $\bar{L}$ could be inside a periodically repeated neighboring unit cell, with $L$ being the corresponding atom in the central cell.
In a similar fashion, the atomic orbitals $\bar{\lambda}[\bar{L}]$ are not restricted to atoms in the central cell, but the  corresponding orbital in the central cell is labeled as $\lambda[L]$.
This enables us to drop the cell index of the real-space overlap matrices, since they are implicitly included for atoms located outside the central cell.

We also follow DFTB+ specific conventions, such that the real-space shift of the overlap $S$ refers to its first index, in line with the notation of Section~\ref{sec:ks-dft-expansion}, and in particular Eq.\,\eqref{eq:Omn-rspace}.
The same reasoning also applies to the real-space density matrix elements.
Let $\lambda$ denote an orbital that is folded back into the central cell from a periodic image, we may then write $S_{\bar{\lambda}\mu} = S_{\lambda\mu}(\bm{h})$, leading to
\begin{align}\label{eq:dHmn-rspace-dftbplus}
\Delta &H^{x,\mathrm{lr}}_{\mu[M]\nu[N]}(\bm{k}) = -\frac{1}{8} \sum\limits_{\substack{\bar{L} \in \mathcal{N}(M) \\ \bar{K} \in \mathcal{N}(N)}} \sum\limits_{\substack{\bar{\lambda}[\bar{L}] \\ \bar{\kappa}[\bar{K}]}} S_{\bar{\lambda}\mu} S_{\bar{\kappa}\nu} \sum_{\bm{g}} \mathrm{e}^{-i \bm{k} \cdot \bm{g}} \Delta P_{\bar{\lambda}\bar{\kappa}}(\bm{g}) \notag
\\
&\times \Big[ \gamma^\mathrm{lr,HF}_{MN}(\bm{g}) + \gamma^\mathrm{lr,HF}_{M\bar{K}}(\bm{g}) + \gamma^\mathrm{lr,HF}_{\bar{L}N}(\bm{g}) + \gamma^\mathrm{lr,HF}_{\bar{L}\bar{K}}(\bm{g}) \Big].
\end{align}
Eq.\,\eqref{eq:dHmn-rspace-dftbplus} exploits the fact that, except in the case of shell-resolved DFTB, the $\gamma$-functions depend only on the atomic species and not the individual orbitals, i.e.\ $\gamma^\mathrm{lr,HF}_{\bar{\lambda}\bar{\kappa}}(\bm{g}) = \gamma^\mathrm{lr,HF}_{\bar{L}\bar{K}}(\bm{g})$ with $\bar{\lambda}[\bar{L}]$ and $\bar{\kappa}[\bar{K}]$.
Analytic expressions for atomic forces, i.e.\ the negative derivative of Eq.\,\eqref{eq:dHmn-rspace-dftbplus} with respect to the ion positions, have been derived and implemented as well.
Although the force expressions for the general $\bm{k}$-point implementation have already been validated against numerical derivatives, their algorithmic optimization is subject to ongoing development and the current implementation is of limited use for production applications.
However, optimized forms of Eq.\,\eqref{eq:dHmn-rspace-dftbplus} for calculations restricted to the $\Gamma$-point, including the energy gradients, are provided in Section~\ref{sec:gamma-point-approx}.

So far, two of the three previously infinite lattice summations have been replaced by well-defined finite summations over the neighbor list, leaving the yet unbounded $\bm{g}$-summation to discuss.

Practical calculations employ a finite set of k-points to sample the first Brillouin zone, with $N_k$ unit cells spanning the BvK supercell.
Restricting the Bloch basis to a finite BvK supercell leads to finite-size errors, since it is not complete with respect to all possible wavevectors, $\bm{k}$, of the infinite crystal.
The density matrix as introduced in Eq.\,\eqref{eq:Pmn-to-rspace} is by construction BvK periodic: $P_{\mu\nu}(\bm{g}) = P_{\mu\nu}(\bm{g} +
\bm{G})$, with $\bm{G}$ denoting a BvK super-lattice vector. This, means it repeats at the boundaries of the BvK supercell without a phase factor. As shown by \citeauthor{hfx-gto}~\cite{hfx-gto}, this artificial periodicity causes lattice sums in the Fock exchange energy expression to diverge.
We have implemented two of the widespread schemes to remedy this issue, namely a truncated Coulomb interaction~\cite{hfx-ct} (TCI) and an adaption of the minimum image convention (MIC)~\cite{hfx-mic}. These enable a robust implementation of periodic Fock exchange.
A more pleasant consequence of the BvK periodicity is that the real-space formulation only requires $N_k$ density matrices, $P_{\mu\nu}(\bm{g})$, to be kept in storage. This is because every lattice shift, $\bm{g}$, may be folded back into the central BvK cell.

\subsection{Truncated Coulomb interaction}\label{sec:tci}
In RSH-DFTB, truncating the Coulomb kernel of the four-center integrals is equivalent to limiting the range of the $\gamma$-integrals
\begin{equation}
\label{eq:truncated-gamma}
\gamma_{MN}^\text{TC}(r) =
\left\{
\begin{array}{ll}
\gamma_{MN}^\mathrm{fr/lr,HF}(R_{MN}) & \quad\text{if } R_{MN} < R_\text{c}
\\
0 & \quad\text{else}
\end{array}
\right.,
\end{equation}
with an adjustable real-space cutoff radius $R_\mathrm{c}$.
To avoid interactions with the neighboring BvK supercells in simple-cubic systems, \citeauthor{hfx-ct}~\cite{hfx-ct} linked $R_\mathrm{c}$ to the number of k-points (which determine the BvK supercell volume).
A more robust scheme for arbitrary lattice geometries, as implemented in this work, is to determine the maximum radius of a sphere that still fits within the BvK supercell~\cite{hfx-gto}.

During self-consistent cycles the geometry and therefore also the $\gamma$-integrals do not change.
Our implementation pre-tabulates all non-vanishing $\gamma_{MN}(R_{MN})$ within the cutoff sphere to speed up
the Hamiltonian construction.

\subsection{Minimum image convention}\label{sec:mic}
Another way of preventing divergent lattice sums was suggested by \citeauthor{hfx-mic}~\cite{hfx-mic} by restricting the sum over
super-lattice vectors according to the minimum image convention.
Later, Irmler and co-workers~\cite{hfx-gto} generalized this to arbitrary k-points.
In this scheme the Coulomb interaction is unaltered and fully taken into account.

We adopt this idea by restricting the argument of the density matrix $\Delta P_{\mu\nu}(\bm{g})$ of Eq.\,\eqref{eq:dHmn-rspace-dftbplus}, such that it does not involve orbitals outside its Wigner-Seitz cell.
This naturally restricts the $\bm{g}$-summation which, depending on the size of the BvK cell, in turn depends on the k-point sampling employed.
To determine the unit cells within the Wigner-Seitz cell of the BvK cell we employ an algorithm that does not assume a specific lattice geometry and works for arbitrary (linearly independent) lattice vectors.

\subsection{Integral pre-screening}\label{sec:screening}
Regardless of whether the TCI or MIC algorithms are used, integral pre-screening targeting the density and overlap matrices has the potential to drastically reduce the computational cost of constructing the exchange Hamiltonian.

In direct self-consistent-field~\cite{direct-scf-1, direct-scf-2} calculations, the Hamiltonian is often constructed iteratively.
Following Lutsker and co-workers~\cite{rsh-dftb-implementation}, the linearity of the Hamiltonian with respect to the density matrix allows representation of the Hamiltonian at the $n$-th self-consistent iteration as a sum of the Hamiltonian at the previous iteration $\Delta {\bf H}(\Delta {\bf P}^{n-1})$ plus a correction $\Delta {\bf H}(\Delta_n (\Delta {\bf P}))$. The change in the density matrix with respect to the previous iteration is denoted as $\Delta_n (\Delta {\bf P})$, resulting in
\begin{align}\label{eq:iterative-hamiltonian}
\Delta {\bf H}(\Delta {\bf P}^n) &= \Delta {\bf H}(\Delta {\bf P}^{n-1} + \underbrace{\Delta {\bf P}^{n} - \Delta {\bf P}^{n-1}}_{\Delta_n (\Delta {\bf P})})
\\
&= \Delta {\bf H}(\Delta {\bf P}^{n-1}) + \Delta {\bf H}(\Delta_n (\Delta {\bf P})).
\end{align}
This approach therefore exploits the rapid decay of $\Delta_n (\Delta {\bf P})$ with increasing cycles of self-consistency.
During the Hamiltonian construction, matrix-matrix products of the form
\begin{align}\label{eq:sps-product}
I_{\bar{L}\bar{K}}^{\mu\nu}(\bm{g}) &:= \sum\limits_{\bar{\lambda}[\bar{L}]} \sum\limits_{\bar{\kappa}[\bar{K}]} S_{\bar{\lambda}\mu} S_{\bar{\kappa}\nu} \Delta_n (\Delta P_{\bar{\lambda}\bar{\kappa}}(\bm{g}))
\end{align}
occur.
The upper bound of Eq.\,\eqref{eq:sps-product} is provided by taking the individual absolute values of the factors
\begin{align}\label{eq:integral-leq}
I_{\bar{L}\bar{K}}^{\mu\nu}(\bm{g}) &\leq \sum\limits_{\bar{\lambda}[\bar{L}]} \sum\limits_{\bar{\kappa}[\bar{K}]} |S_{\bar{\lambda}\mu}| |S_{\bar{\kappa}\nu}| |\Delta_n (\Delta P_{\bar{\lambda}\bar{\kappa}}(\bm{g}))| \notag
\\
&\leq S^\mathrm{max}_{\bar{L}M} S^\mathrm{max}_{\bar{K}N} \Delta_n (\Delta {\bf P})^\mathrm{max} \sum\limits_{\bar{\lambda}[\bar{L}]} \sum\limits_{\bar{\kappa}[\bar{K}]} 1,
\end{align}
where maximum estimates for the overlap $S^\mathrm{max}_{\bar{L}M} := \max_{\bar{\lambda}[\bar{L}], \mu[M]}(|S_{\bar{\lambda}\mu}|)$ and density matrix $\Delta_n (\Delta {\bf P})^\mathrm{max} := \max_{\bm{g}}(|\Delta_n (\Delta {\bf P}(\bm{g}))|)$ have been defined.
If $S^\mathrm{max}_{\bar{L}M} S^\mathrm{max}_{\bar{K}N} \Delta_n (\Delta {\bf P})^\mathrm{max} < \varepsilon_\mathrm{screen}$, where the trivial summations over orbitals $\bar{\lambda}, \bar{\kappa}$ have been absorbed by the integral screening parameter $\varepsilon_\mathrm{screen}$, the evaluation of the corresponding diatomic sub-block of $\Delta {\bf H}(\Delta_n (\Delta {\bf P}))$ is omitted.
In preparation for the evaluation of Eq.\,\eqref{eq:dHmn-rspace-dftbplus}, all occurring ${\bf S}(\Delta {\bf P}){\bf S}$-products are estimated and the terms requiring an explicit evaluation are distributed to available processors (provided that a message passing interface (MPI) parallelized version of DFTB+ is being used).

\subsection{Scaling with system size}\label{sec:scaling}
\begin{figure}[htbp]
\centering
\includegraphics[width=1.0\columnwidth]{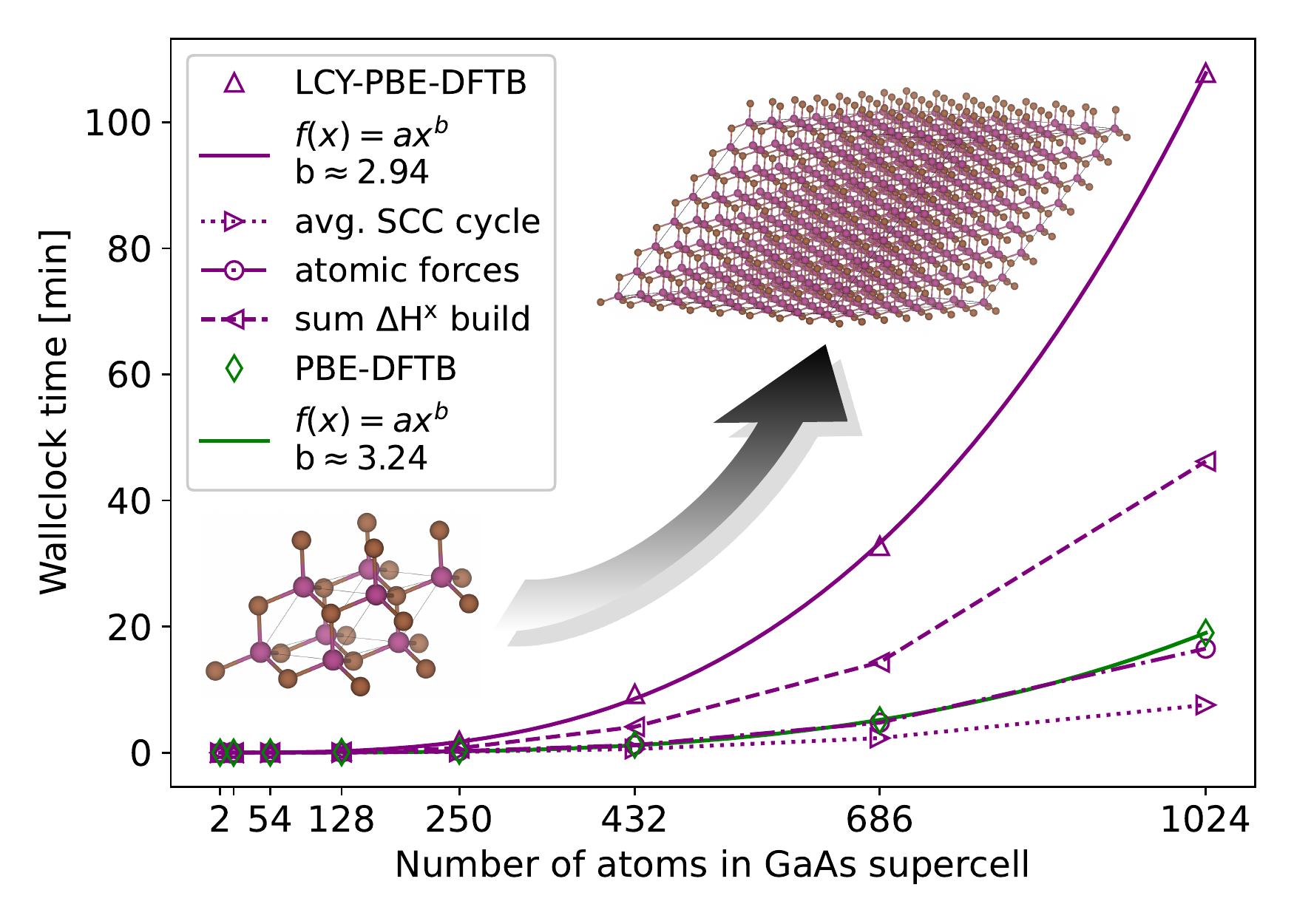}
\caption{\label{fig:benchmark-gamma_implementation} Total wall-clock time of a RSH-DFTB $\Gamma$-point calculation with the LCY-PBE xc-functional, in comparison with conventional PBE-DFTB, as performed on a single CPU core (AMD Ryzen 7 PRO 5850U). For LCY-PBE-DFTB, the total time spend on an average self-consistent cycle, the construction of the RSH contribution to the total Hamiltonian and atomic force evaluation is indicated as well. GaAs supercells are employed as model system. LCY-PBE-DFTB exhibits slightly sub-cubic scaling, which might indicate that the asymptotic limit is not yet reached.}
\end{figure}

While supercells of several hundreds of atoms are often unattainable for proper long-range corrected hybrid functionals within RSH-DFT, we
demonstrate that such cases are well within reach of RSH-DFTB, even on a single processor core and for relatively densely packed materials such as GaAs.
Figure~\ref{fig:benchmark-gamma_implementation} compares the total wall-clock time of a RSH-DFTB $\Gamma$-point calculation using the LCY-PBE functional and Yukawa type range-separation function, to a conventional PBE-parameterized DFTB (referred to as PBE-DFTB) run.
GaAs is computationally challenging due to its large number of interacting neighbors.
The computational cost of LCY-PBE-DFTB turns out to be considerably higher than for conventional PBE-DFTB.
However, considering that the benchmark was performed on a single CPU core only, calculations of large supercells with roughly 1000 atoms can be accomplished in reasonable time.
The higher cost of LCY-PBE-DFTB mainly originates from three facts: a) the pre-tabulation of all $\tilde{\gamma}_{MN}^\mathrm{TC}$ to build the super-matrix $\tilde{\gamma}^\mathrm{TC}$ (see Section~\ref{sec:gamma-point-approx}), b) the actual time spend on constructing the Hamiltonian of Eq.\,\eqref{eq:dHmn-rspace-gamma-tc-mm} and c) a higher number of total self-consistency steps, compared to traditional DFTB.
Reason c) is expected, since (semi-)local DFTB requires only self-consistency with respect to the Mulliken populations, while RSH-DFTB introduces terms that depend on the full density matrix. Its self-consistency is with respect to the (real-space) matrices $\Delta P_{\mu\nu}(\bm{g})$.
In other words, traditional DFTB mixes the in- and out-put Mulliken populations to propagate the self-consistent cycles, while RSH-DFTB mixes $\Delta P_{\mu\nu}(\bm{g})$, which proves to be more challenging and leads in most cases to an increased number of self-consistency steps, and therefore diagonalizations of the total Hamiltonian.

\subsection{Parallel performance}\label{sec:mpi}
\begin{figure}[htbp]
\centering
\includegraphics[width=1.0\columnwidth]{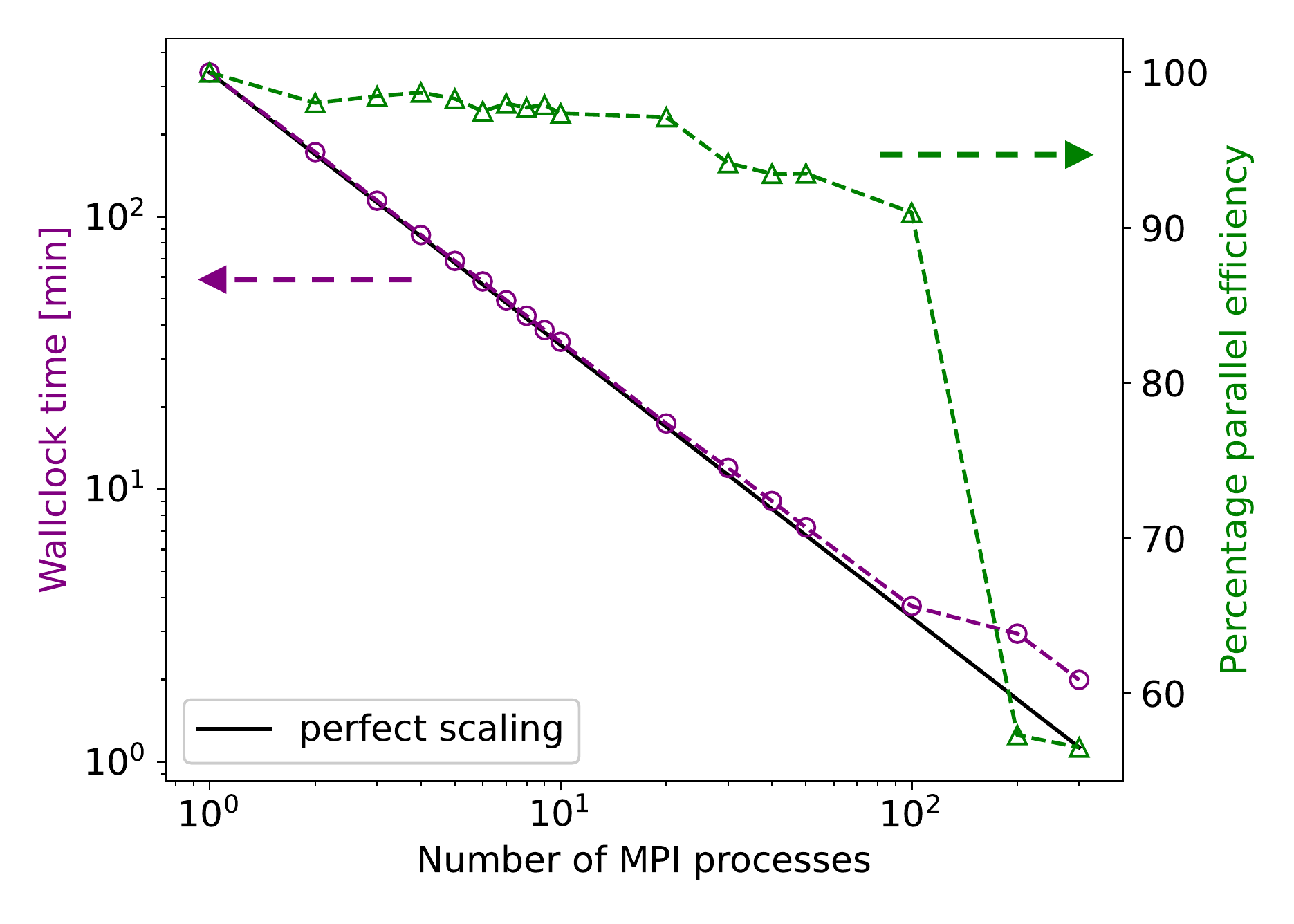}
\caption{\label{fig:benchmark-MPI_parallelism} Parallel performance of DFTB+, when performing range-separated calculations beyond the $\Gamma$-point ($\varepsilon_\mathrm{screen}$ is set at $10^{-7}\,\mathrm{a.u.}$). The primitive GaAs unit cell, sampled by a $9\times 9\times 9$ Monkhorst-Pack k-point set, served as model system to obtain the (I/O time removed) wall-clock time of the entire DFTB+ run (with one MPI process corresponding to one processor core). Up to about 100 cores the parallel efficiency is excellent, however a slight change in slope due to incipient inter-node communication between 20 and 30 cores is observed. The parallel efficiency then decreases to about $50-60\,\%$ when further increasing the core count. Considering the extremely small test system of only two atoms, this saturation is expected. The employed HPC provides nodes of two Intel Xeon E5-2690v4 CPUs (2.6 GHz, 14 cores each), resulting in 28 cores per node, whereas inter-node communication is based on Intel's Omni-Path network architecture.}
\end{figure}

The relatively high cost of constructing the exchange Hamiltonian in RSH-DFTB requires an efficient parallelization of this step, in order to exploit modern computing infrastructures and HPC facilities.
While Section~\ref{sec:scaling} already demonstrated the suitability for large supercells, another common task is the calculation of smaller systems with a dense k-point sampling.
Figure~\ref{fig:benchmark-MPI_parallelism} illustrates the parallel performance for the energy evaluation of a primitive GaAs unit cell.
The scaling of the total wall-clock time with MPI processes is quite satisfactory and only saturates from about 100 cores onwards, for this system.
One bottleneck that causes the parallel efficiency to drop when exceeding this core count, at least in the current implementation, concerns the mixing of input and output density matrices in the self-consistency loop of DFTB+, which is not yet MPI parallelized.
This also affects the memory consumption.
For some mixers (e.g.\ modified Broyden's method~\cite{broyden-mixer}) the history of all previous $\Delta P_{\mu\nu}(\bm{g})$ is kept in storage, which becomes unfeasible for extremely dense k-point samplings or an unusual large number of self-consistent steps to reach convergence.
Fortunately, mixing schemes with limited memory (e.g.\ modified Anderson's method~\cite{anderson-mixer}) are readily available within DFTB+.

\section{Convergence behavior}\label{sec:convergence}
\subsection{Polyacene series}\label{sec:acene}
The important class of $\pi$-conjugated polymers has spawned numerous successful candidates for devices like organic light-emitting diodes (OLEDs)~\cite{oled-1, oled-2, oled-3}, organic field-effect transistors (OFETs)~\cite{fet-1, fet-2, fet-3}, polymer solar cells (PSCs)~\cite{psc-1, psc-2, psc-3} and the growing field of organic electronics in general.
One representative of this class is the \cf{C_{4n+2}H_{2n+4}} series, forming the polyacene oligomers.
We choose this linear molecular chain due to its relevance as a previous benchmark system for range-separated DFT.
For many $\pi$-conjugated polymers (semi-)local DFT fails to describe the bond length alternation (BLA) and band-gap correctly.
While Hartree-Fock (HF) overestimates BLA significantly, (semi-)local DFT is known to underestimate it.
In fact, \citeauthor{bla-sie}~\cite{bla-sie} suggested that the many-electron self-interaction error (MSIE) of HF and DFT approaches
correlate with the BLA error and MSIE minimization is key (but not the only issue) to obtaining accurate BLAs.
These quasi one-dimensional systems with low environmental screening also provide a stringent test for the removal of the divergence in the exchange interaction. In addition, the polyacenes feature a well-defined finite molecular limit, which can be used to verify the periodic RSH-DFTB implementation proposed here.

\begin{figure}[htbp]
\centering
\includegraphics[width=1.0\columnwidth]{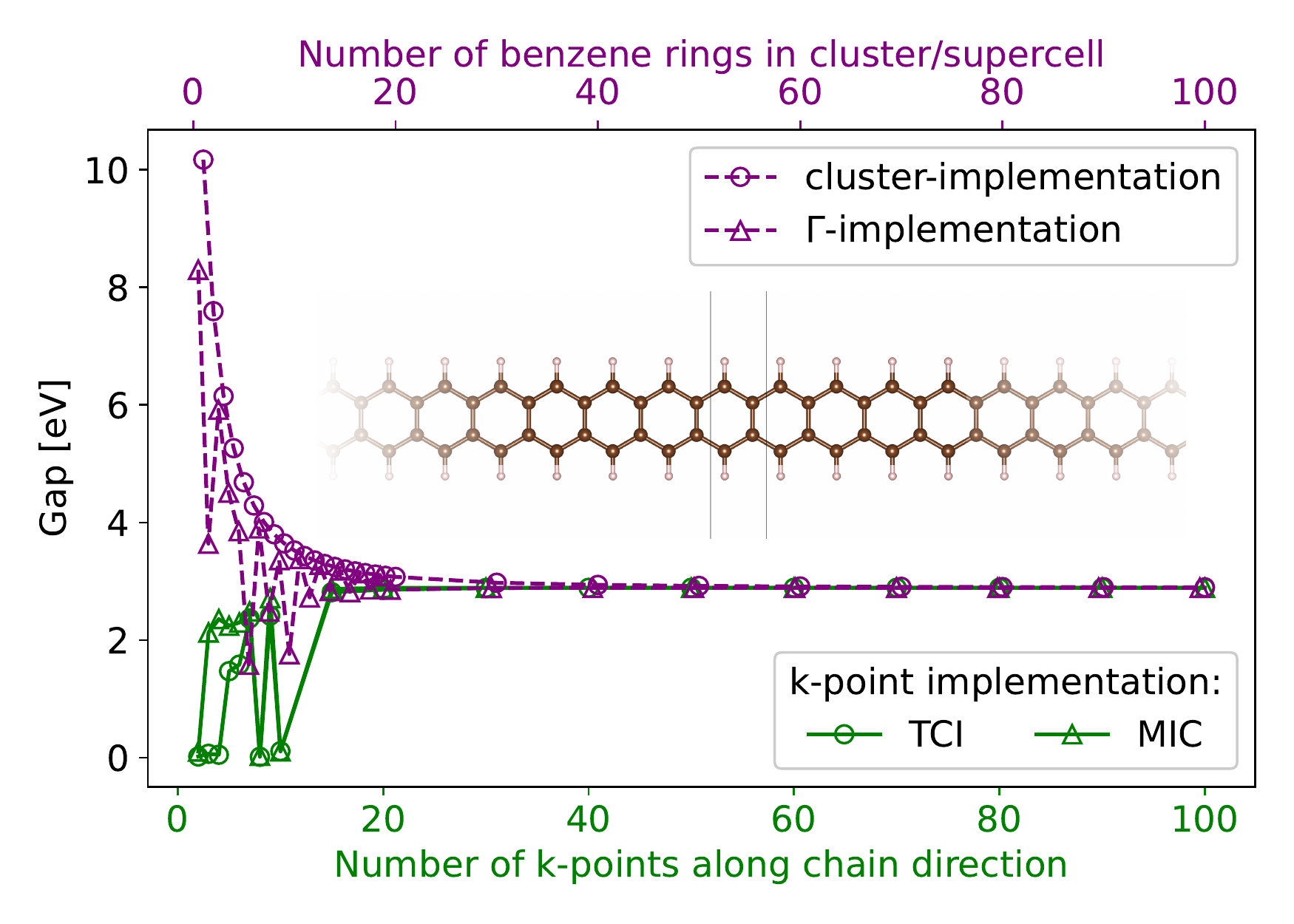}
\caption{\label{fig:polyacene} Band-gap convergence behavior of the periodic implementations ($\Gamma$-point, TCI, MIC) for the polyacene series, in comparison with the non-periodic algorithm. In the limit of dense k-points or large clusters/supercells, respectively, all implementations agree ($E_\mathrm{gap} = 2.9\,\mathrm{eV}$).}
\end{figure}

Since (semi-)local DFTB is directly derived from DFT, it also inherits its shortcomings.
A particularly severe deficiency of PBE-DFT(B) is that the polyacene series becomes metallic for increasing chain length~\cite{Niehaus2005a}.
We employ the ob2-1-1 parameters~\cite{ob2-params}, created for the purely long-range corrected LCY-BNL functional, to demonstrate that our implementation of periodic RSH-DFTB converges to the same, finite band-gap as the already available non-periodic formalism.
We provide an estimate of the polyacene bandgap for the family of purely long-range corrected functionals, as calculated by the FHIaims code~\cite{fhiaims, fhiaims-1, fhiaims-2} on the LC-$\omega$PBE~\cite{LC-wPBE} level of theory using intermediate basis settings and a range-separation parameter of $\omega = 0.3\,{a_0^{-1}}$, yielding $E_\mathrm{gap} = 2.8\,\mathrm{eV}$~\footnote{Due to technical difficulties experienced with the latest release version 221103 of the FHIaims code, this calculation has been carried out with version 210716.3 instead.}.
Since the specific LC-BNL functional is not yet available through FHIaims (release version 221103)~\footnote{FHIaims in version 221103 is interfaced with a recent version of the libXC library (5.1.7), that, in principle, provides an implementation of the BNL functional. However, LDA based hybrid functionals like BNL do not seem to be supported by FHIaims in version 221103. Furthermore, to the best of our knowledge, a manual adaption of the range-separation parameter $\omega$ is not yet supported, including the current development version of FHIaims, rendering a quantitative comparison of the ob2-1-1 parameters with the literature parametrization of the BNL functional~\cite{rsh2}, with $\omega = 1\,a_0^{-1}$, meaningless.}, this value does not allow for a quantitative comparison with results obtained by the ob2-1-1 parameters, however, offers valuable guidance from first principles.
Figure~\ref{fig:polyacene} illustrates the convergence of the polyacene band-gap convergence for the present periodic $\Gamma$-point and k-point implementation, in direct comparison to the non-periodic case.
The calculations are based on the (unrelaxed) primitive unit cell of polyacene (k-point implementation), as listed in Structure~S1 of the Supplemental Material~\cite{supplmat}, and supercells (the $\Gamma$-point implementation) built from it.
In the case of the non-periodic implementation, the supercells are converted into clusters and properly passivated by additional hydrogen atoms at the chain ends (bond-length of C-H units: $1.1\,\text{\AA}$).
The k-point sampling $1\!\times\! 1\!\times\! (2 \leq n \leq 100)$ is chosen according to the Monkhorst-Pack~\cite{mp-scheme} scheme, where the polyacene chain is oriented along the $z$-direction and vacuum inserted in $x$- and $y$-direction.
What immediately stands out is that all implementations converge towards the same Kohn-Sham gap, which is an essential step in the validation~\footnote{Another strategy for verifying the correct implementation that has proven to be useful, is to test the translational invariance of the method by homogeneously shifting the coordinates (partly outside the unit cell) and verifying that the Kohn-Sham spectrum is unaltered.} of the present method.
A closer look at the convergence behavior of the general k-point implementations of the TCI and MIC schemes reveals strong fluctuations of the band-gap for non-convergent k-point samplings $1\!\times\! 1\!\times n$, where $n \lesssim 20$.
These fluctuations originate from artifacts in the band-structure and can be avoided by either manually reducing the Coulomb truncation cutoff in TCI or removing outer shells of unit cells inside the Wigner-Seitz cell of the BvK supercell in MIC.
We take the opportunity of this specific case to highlight the pitfalls of periodic Fock exchange in RSH-DFT(B).
\begin{figure}[h!]
\centering
\includegraphics[width=1.0\columnwidth]{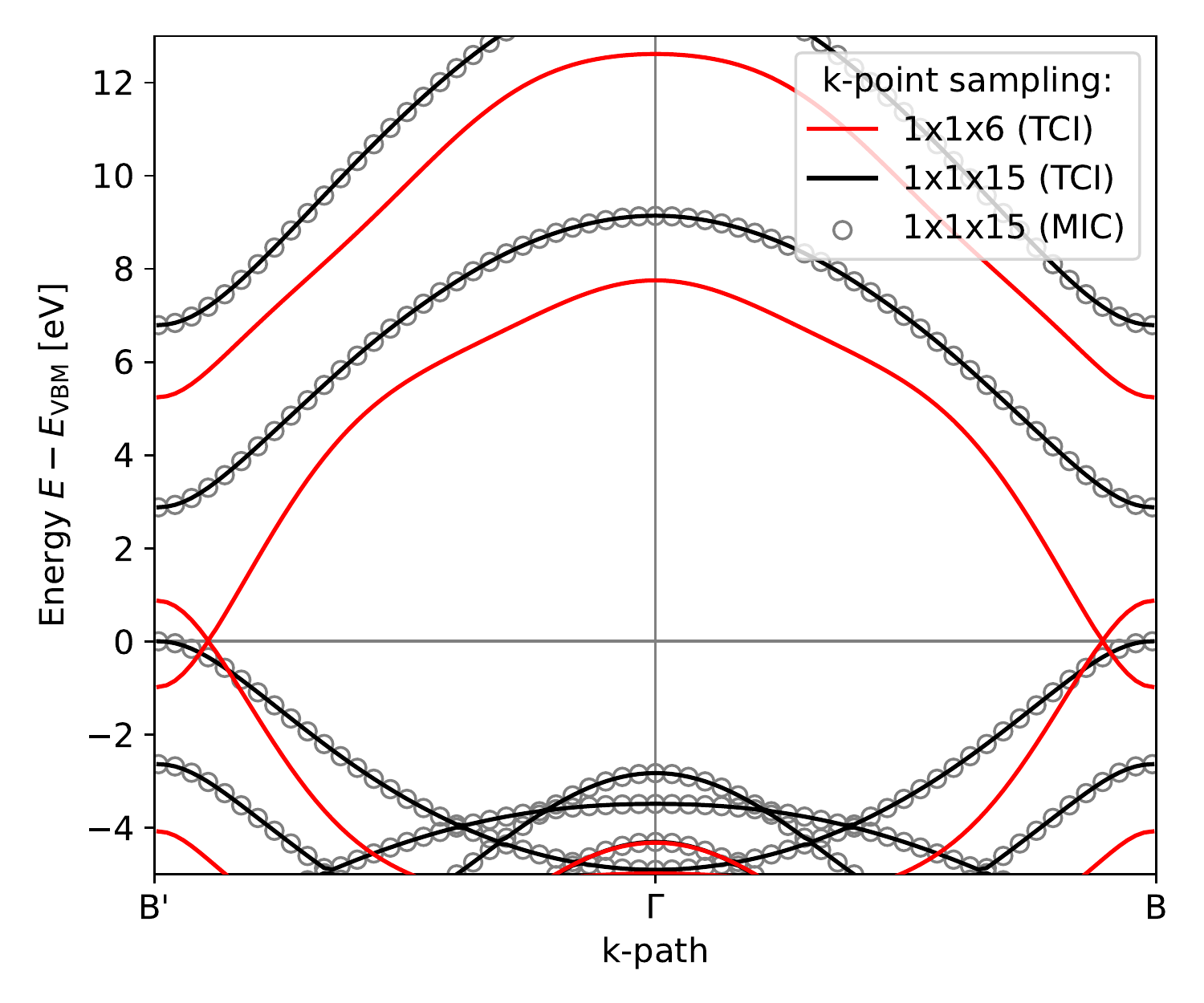}
\caption{\label{fig:acene-bandstructures} Band-structures of the primitive polyacene unit cell, aligned at their respective valence band maximum (VBM). Fully converged RSH-DFTB calculations, with $1\!\times\! 1\!\times\! 15$ k-point sampling in the self-consistent run that produced the ground-state density, performed in the TCI and MIC scheme, are compared to a band-structure that exhibits artifacts due to an insufficient truncation of the Coulomb interaction for small BvK supercells. The first Brillouin zone was sampled between $B' = (0.0, 0.0, -0.5)$ and $B = (0.0, 0.0, +0.5)$.}
\end{figure}
Figure~\ref{fig:acene-bandstructures} explicitly shows two fully converged band-structures of Figure~\ref{fig:polyacene}, demonstrating that not only the gap size, but also all bands, calculated with TCI or MIC are virtually identical.
Additionally, a non-convergent band-structure, originating from a density calculation with $1\!\times\! 1\!\times 6$ k-point sampling is included.
For this choice of parameters the band-gap collapses and individual bands exhibit an unphysical dispersion.
Too small a BvK supercell does not allow for a natural decay of the density matrix, but rather introduces a spurious periodicity as described in Section~\ref{sec:neigh-algo}.
The extent to which artifacts of non-convergent calculations based on coarse k-point samplings manifest themselves is system-specific.
As an example shown in Figure~S1 of the Supplemental Material~\cite{supplmat}, armchair graphene nanoribbons (AGNRs) turn out to be a much more benign system and convergence is achieved rapidly.
We refer to Structures~S3-S5 and Figures~S2-S3 of the Supplemental Material~\cite{supplmat} to obtain further investigations covering two-dimensional h-BN monolayer and GaAs bulk.

\subsection{Total energy convergence}\label{sec:energy-convergence}
The total energy is often considered to provide a solid indication of the convergence behavior of a system.
In the literature~\cite{algo-ct, hfx-gto, hfx-ct} it is used for the sake of comparing implementations and to demonstrate convergence.
However, this measure proved to be unreliable in many cases with regard to other properties,
including band-gaps.
We would therefore like to point out that, at least for periodic RSH-DFTB, further quantities of interest should also be considered when checking convergence.
\begin{figure}[h!]
\centering
\includegraphics[width=1.0\columnwidth]{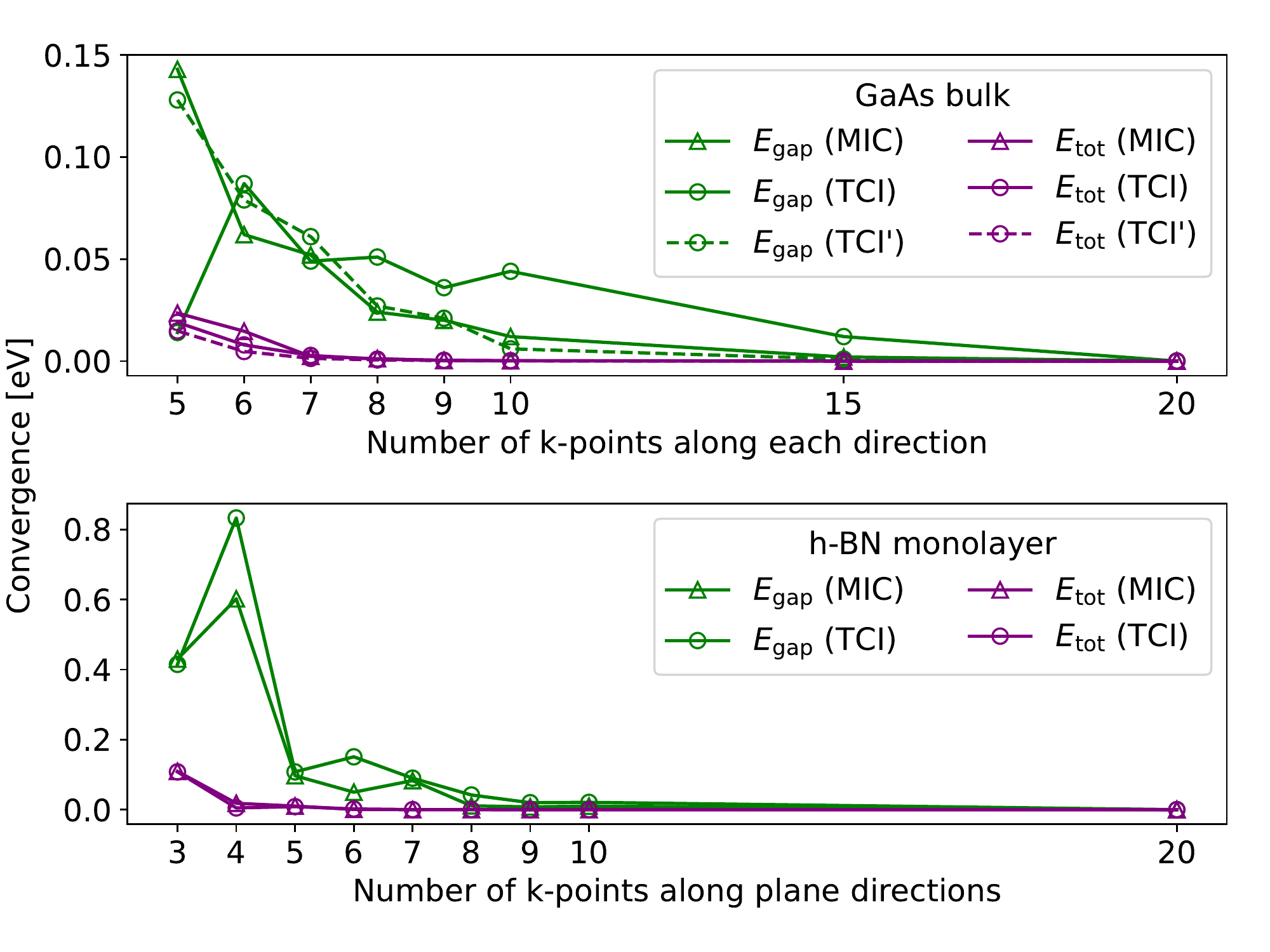}
\caption{\label{fig:energy-vs-property_convergence} Absolute deviation in the total energy and band-gap with respect to their fully converged reference. For periodic RSH-DFTB calculations, the total energy of a system is empirically found to converge more rapidly than the Kohn-Sham band-gap. For GaAs bulk it was found that setting the TCI cutoff as half of the minimum lattice vector norm ($\mathrm{TCI}'$), the total energy and band-gap converge more quickly when compared to the scheme described in Section~\ref{sec:tci}. The choice of coupling the TCI cutoff to the k-point sampling can therefore significantly influence the convergence behavior of a system.}
\end{figure}
Figure~\ref{fig:energy-vs-property_convergence} illustrates that the absolute deviation in the total energy and band-gap with respect to their fully
converged reference decreases at different rates, with slower convergence of the band-gap.
This phenomenon appears to be independent of the dimensionality of the system and applies, e.g., for three-dimensional GaAs bulk and two-dimensional h-BN monolayer included in Figure~\ref{fig:energy-vs-property_convergence}.
To obtain a fully converged total energy and band-structure as reference, we employed a GaAs density calculation with a $20\times\! 20\times\!
20$ and h-BN monolayer calculation with $20\times\! 20\times\! 1$ Monkhorst-Pack k-point sampling respectively.

\section{Polarization-induced gap renormalization}\label{sec:gaprenorm}
Renormalization of the fundamental band-gap in molecular crystals by electronic polarization~\cite{gap-renorm-first} is of central importance for organic electronics~\cite{oled-1, oled-2, oled-3, fet-1, fet-2, fet-3} and photovoltaics~\cite{psc-1, psc-2, psc-3}.
Going from the molecule in gas phase to a molecular crystal with relative dielectric constant $\varepsilon_\infty$ (orientationally
averaged and ion-clamped) leads to shrinkage of the fundamental gap.
This renders the resulting material well-suited for practical applications that require reduced optical gaps.
Due to the electronic polarization of the crystalline dielectric medium, the energy required to create a quasi-hole is reduced compared to its molecular phase, whereas creating a quasi-electron releases more energy~\cite{gap-renorm}.
In other words, the ionization potential (IP) and electron affinity (EA) decrease and increase respectively.

Today's standard repertoire of exchange-correlation functionals within DFT, including (semi-)local LDA/GGA as well as global and
range-separated hybrids~\cite{rsh}, do not properly treat long-range correlation effects and fail to describe the aforementioned gap renormalization, even qualitatively~\cite{dft-no-epsilon}.
While many-body perturbation theory, especially Hedin's GW approximation~\cite{gw} to the electron's self-energy, $\Sigma$,
captures these renormalization effects, only recent screened range-separated hybrid functionals~\cite{gap-renorm} include this effect at the considerably cheaper level of DFT.

\begin{figure*}[htbp]
\centering
\includegraphics[width=0.9\textwidth]{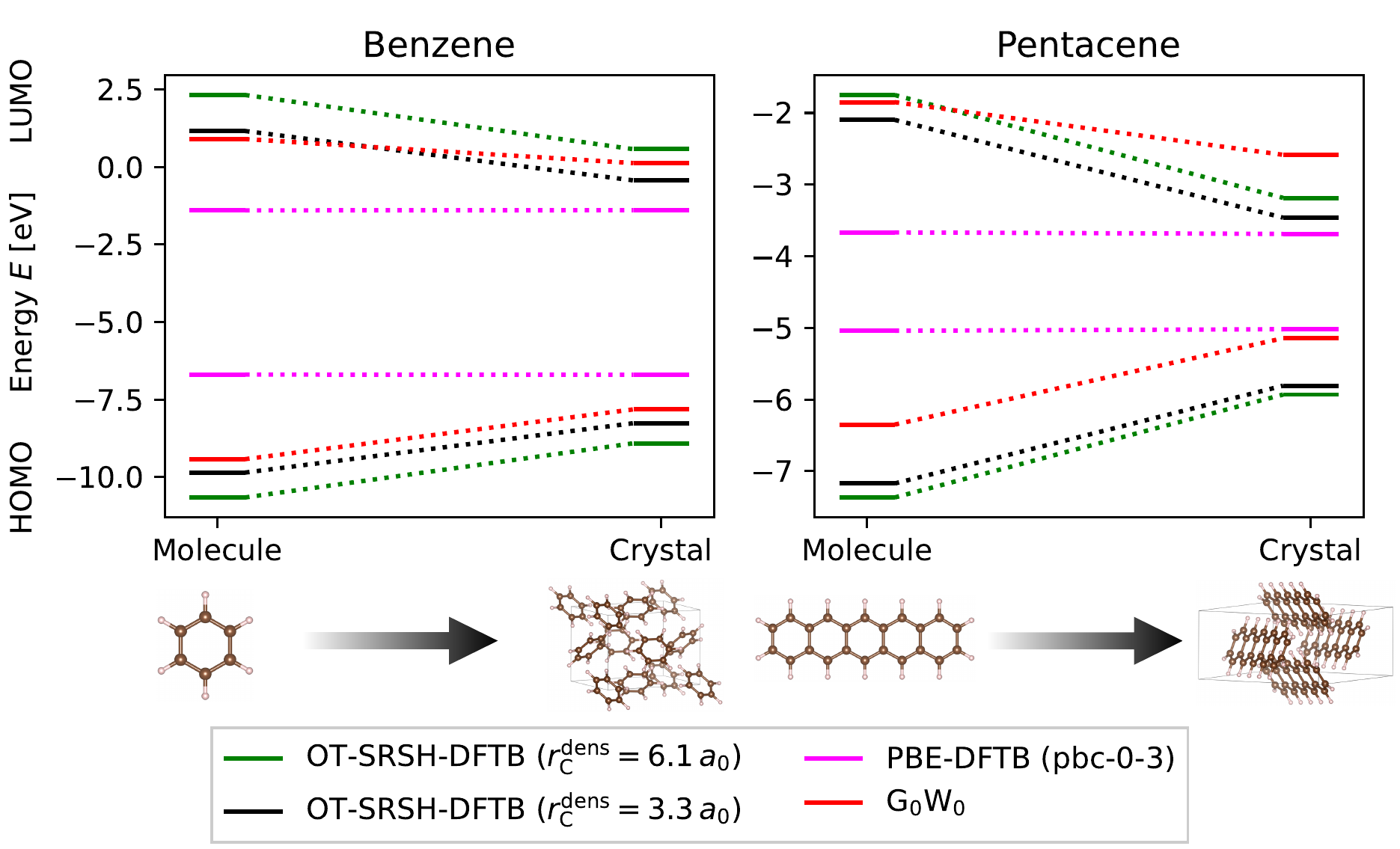}
\caption{\label{fig:gap-renorm} Fundamental gaps of gaseous and crystalline benzene and pentacene, as obtained by conventional PBE-DFTB and optimally tuned screened range-separated hybrid functionals (OT-SRSH-DFTB), compared with $\mathrm{G_0W_0}$ calculations. To maintain comparability with Ref.~\onlinecite{gap-renorm}, crystalline gaps were aligned to the middle of their respective gas-phase gap. $\mathrm{G_0W_0}$ values were taken from Refs.~\onlinecite{g0w0-benzene-gas, g0w0-pentacene-gas} (gas-phase) and Ref.~\onlinecite{gap-renorm} (bulk).}
\end{figure*}

According to the general CAM partitioning of the electron-electron interaction introduced by Eq.\,\eqref{eq:cam}, the limiting behavior of the long-range part, that we treat in an exact Fock-like manner, is determined by the parameters $\alpha$ and $\beta$.
For small distances as $r\to 0$, the $\alpha / r$ contribution prevails, whereas the limiting case of $r\to\infty$ scales like $(\alpha + \beta) / r$.
In the gas phase, the correct $1 / r$ asymptotic decay is obtained if the condition $\alpha + \beta = 1$ is fulfilled.
In fact this only represents the special case of $\varepsilon_\infty = 1$ and a generalization to arbitrary dielectric environments with asymptotic potential of $1/(\varepsilon_\infty r)$ requires that $\alpha + \beta = 1/\varepsilon_\infty$.
For the gas-phase fundamental gap to coincide with the HOMO-LUMO gap of generalized Kohn-Sham RSH-DFT~\cite{ea-ip-1, ea-ip-2}, the
Fock-like exchange term is required to be asymptotically correct~\cite{fgap-HL-1, g0w0-pentacene-gas} and the range-separation parameter
$\omega$ should be tuned to obey the ionization-potential (IP) theorem~\cite{piecewise-linear, ip-theorem-2, ip-theorem-3, ip-theorem-4}.
We follow~\citeauthor{gap-renorm}~\cite{gap-renorm} by non-empirically determining $\omega$, through minimizing the function
\begin{align}
J^2(\omega ; \alpha) &= \left( \varepsilon_\mathrm{H,n}^{\omega, \alpha} + \mathrm{IP}_\mathrm{n}^{\omega, \alpha} \right)^2 + \left( \varepsilon_\mathrm{H,a}^{\omega, \alpha} + \mathrm{IP}_\mathrm{a}^{\omega, \alpha} \right)^2
\end{align}
for the gas-phase. Here $\varepsilon_\mathrm{H,n}^{\omega, \alpha},\varepsilon_\mathrm{H,a}^{\omega, \alpha}$ denote the energies associated with the HOMO of the neutral (n) and anionic (a) systems, tuned to match the respective IP obtained from total energy differences as closely as possible.
Additional figures outlining the optimization process are provided in Section~S4 of the Supplemental Material~\cite{supplmat}.
For the sake of comparing our results with Ref.~\onlinecite{gap-renorm}, we do not attempt to optimize $\alpha$ from first principles, but rather chose $\alpha = 0.2$, which proved to yield satisfactory results for small organic molecules.
The employed exchange-correlation functional is CAMY-PBEh~\cite{cam-pbeh} (with Y indicating the range-separation function is of Yukawa type).
In order to consistently compare the results obtained with our RSH-DFTB method, we resort to the molecular and crystalline geometries of Ref.~\onlinecite{gap-renorm}, covering the prototypical conjugated molecules benzene and pentacene.
The same reasoning applies to the choice of the scalar dielectric
constant, which is also taken from the cited reference.
A full overview of the resulting functional parameterization, including the optimally tuned range-separation parameters $\omega$, is provided in Table~S2 of the Supplemental Material~\cite{supplmat}.
To avoid time-consuming re-parameterization, the generated Slater-Koster files are based on the ob2-1-1~\cite{ob2-params} parameters, which are expected to perform well in combination with long-range corrected functionals.
Additionally, results obtained by a slight modification to the ob2-1-1 parameters are shown, this decreases the density compression radius of the carbon species to a value of $r^\mathrm{dens}_\mathrm{C} = 3.3\,a_0$.
We refer to Table~S1 of the Supplemental Material~\cite{supplmat} for a detailed listing of the electronic parameterization.
Figure~\ref{fig:gap-renorm} compares the fundamental gaps obtained using different levels of theory.
The PBE-DFTB calculations were carried out based on the pbc-0-3 parameters~\cite{pbc-params}.
Since the hydrogen and carbon electronic structure of pbc-0-3 is virtually identical to mio-1-1~\cite{mio-params}---the usual
choice for biological or organic molecules within second-order DFTB---only one of the two options is included for comparison.
In line with expectations, PBE-DFTB is unable to capture changes in the surrounding dielectric medium and severely underestimates not only the molecular but even the crystalline fundamental gap.
In stark contrast, optimally tuned screened range-separated hybrid (OT-SRSH) DFTB as derived and implemented in this work captures the renormalization phenomenon at least qualitatively.
However, the gas-phase HOMO is slightly too low in both systems, which also affects the bulk phase due to the chosen crystalline gap
alignment.
To maintain comparability with Ref.~\onlinecite{gap-renorm}, the crystalline gap was aligned symmetrically around the middle of the respective gas-phase gap.
Quantitative agreement would require a more in depth re-parameterization, which is outside the scope of this work.
We emphasize that OT-SRSH-DFTB is systematically extending the domain of accessible exchange-correlation functionals within (periodic) DFTB
by only introducing system-specific adjustable parameters. These are determined from first principles in a well-founded tuning process, rather than being subject to empirical fitting.

\section{Summary and outlook}\label{sec:s&o}
We have derived, implemented and tested Hartree-Fock exchange in the density functional tight-binding (DFTB) method for periodic systems beyond the $\Gamma$-point.
By applying the usual DFTB approximations to matrix elements of the Fock exchange operator, we arrived at a real-space formulation of the
exchange Hamiltonian that allows for efficient implementation in the DFTB+ code.
To avoid artifacts owing to the artificial Born--von K\'arm\'an periodicity of the density matrix, we resort to either a truncated Coulomb
interaction or a minimum image convention.
These two methods proved to converge to the same limit, when coupled to accurate Brillouin zone samplings.
Scaling with system size and parallel performance of the developed routines indicate that systems with thousands of atoms are well within
reach of RSH-DFTB and that computational resources are exploited efficiently.
Pre-screening of products of density matrix and overlaps, in combination with an iterative construction of the Hamiltonian allows us to further reduce the computational cost.
Convergence behavior and the pitfalls of periodic RSH-DFTB are demonstrated for the polyacene series, showing that periodic and non-periodic implementations converge towards the same limits, and that the total energy is often not a reliable indicator for the convergence
of other properties like the band-gap.
In line with $G_0 W_0$ calculations, screened RSH-DFTB shows the correct polarization-induced gap renormalization in benzene and pentacene molecular crystals, at a heavily reduced computational cost compared to first principles methods.

An in-depth benchmarking of the accuracy of periodic RSH-DFTB is currently subject of ongoing investigations and will be the topic of a future article.

\begin{acknowledgments}
T.\ v.\,d.\,H.\  and T.\ A.\ N.\ acknowledge financial support from the German Research Foundation (DFG) through Grant No.\ FR2833/76-1.
The simulations were performed on the HPC cluster \emph{Aether} at the University of Bremen, financed by the German Research Foundation (DFG) within the scope of Zukunftskonzept 66 “Ambitioniert und agil”, Bremen (GZ ZUK 66/1-2015).
\end{acknowledgments}

\section*{Data availability}
The data that supports the findings of this study are available within the article and its Supplemental Material~\cite{supplmat}.

\appendix
\section{Hubbard parameter of RSH-DFTB}\label{sec:hubbu-cam-dftb}
In analogy to conventional DFTB, the Hubbard $U$ of RSH-DFTB is derived by requiring its equality with RSH-DFT.
Further, the xc-kernel kernel $f_\mathrm{xc,loc}$ is assumed to vanish~\cite{rsh-dftb-implementation} for off-site elements $A\neq B$, whereas the on-site elements $A=B$ cover the full exchange-correlation contributions and read as
\begin{align}
\lim_{R_{AA} \to 0} \gamma_{AA}^\mathrm{fr}(R_{AA}) &= \frac{5}{16} \tau_A \label{eq:gamma-fr-zero-limit}
\\
\lim_{R_{AA} \to 0} \gamma_{AA}^\mathrm{lr,HF}(R_{AA}) &= \frac{5}{16} \tau_A - \frac{\tau_A^8}{(\tau_A^2 - \omega^2)^4} \notag
\\
&\times \left[ \frac{5\tau_A^6 + 15\tau_A^4\omega^2 - 5\tau_A^2\omega^4 + \omega^6}{16\tau_A^5} - \omega \right].
\end{align}
Note that, by definition, the function $\gamma^\mathrm{fr,HF} = \lim_{\omega \to \infty} \gamma^\mathrm{lr,HF}$ is always contained in the results for the more general screened Coulomb kernel.
Since the Hubbard $U$ is obtained from a single atom RSH-DFT calculation, the following conditions apply
\begin{align}
S_{\mu\nu} &= \delta_{\mu\nu} \label{eq:hubbard-condition-1}
\\
\gamma_{\mu\nu}^\mathrm{fr,HF} &= \gamma_{AA}^\mathrm{fr,HF}
\\
H^{(0)}_{\mu\nu} &= \delta_{\mu\nu}\varepsilon^\mathrm{free}
\\
\sum_{\mu} c_{\mu i}c_{\mu j} &= \delta_{ij}. \label{eq:hubbard-condition-4}
\end{align}
By utilizing Eqs.\,\eqref{eq:hubbard-condition-1} to \eqref{eq:hubbard-condition-4}, the total RSH-DFTB energy of the single atom in terms of occupation numbers $n_i$ is~\cite{rsh-dftb-implementation}
\begin{align}
E^\mathrm{atom} &= \sum_{\mu} P_{\mu\mu} \varepsilon^\mathrm{free}_\mu + \frac{1}{2} \gamma_{AA}^\mathrm{fr} \sum_{\mu\kappa} \Delta P_{\mu\mu} \Delta P_{\kappa\kappa} \notag
\\
&- \frac{1}{4} \left( \alpha \gamma_{AA}^\mathrm{fr,HF} + \beta \gamma_{AA}^\mathrm{lr,HF} \right) \sum_{\mu\nu} \Delta P_{\mu\nu} \Delta P_{\mu\nu}
\\
&= \frac{1}{2} \sum_{ij} n_i n_j \gamma_{AA}^\mathrm{fr} - \frac{1}{4} \sum_i n_i^2 \left( \alpha \gamma_{AA}^\mathrm{fr,HF} + \beta \gamma_{AA}^\mathrm{lr,HF} \right) + \mathcal{O}(n_i), \label{eq:Eatom-ni}
\end{align}
with terms linear in $n_i$ indicated by $\mathcal{O}(n_i)$.
For the highest occupied shell it holds that for $n$ electrons, equally distributed over the shell, give an orbital occupation $n_i = n/d_l$, where the shell degeneracy is $d_l = 2l + 1$.
Inserting $d_l$ into Eq.\,\eqref{eq:Eatom-ni}, yields
\begin{align}\label{eq:Eatom-n}
E^\mathrm{atom} &= \frac{1}{2} \gamma_{AA}^\mathrm{fr} n^2 - \frac{1}{4} \left( \alpha \gamma_{AA}^\mathrm{fr,HF} + \beta \gamma_{AA}^\mathrm{lr,HF} \right) \frac{n^2}{d_l} + \mathcal{O}(n).
\end{align}
Calculating the second derivative of Eq.\,\eqref{eq:Eatom-n} with respect to the shell occupation $n$ then becomes straightforward:
\begin{align}\label{eq:Eatom-sec-deriv}
\frac{\partial^2 E^\mathrm{atom}}{\partial n^2} &= \gamma_{AA}^\mathrm{fr} - \frac{1}{2} \frac{1}{2l + 1} \left( \alpha \gamma_{AA}^\mathrm{fr,HF} + \beta \gamma_{AA}^\mathrm{lr,HF} \right).
\end{align}
In practice, the Hubbard $U$ of RSH-DFT is obtained by numerically calculating the second derivative of the eigenvalue $\varepsilon_\mathrm{HOAO}^A$ of the highest occupied atomic orbital of species $A$ with respect to its occupation $n_\mathrm{HOAO}$
\begin{align}
U^\mathrm{RSH-DFT}_A &= \frac{\partial\varepsilon_\mathrm{HOAO}^A}{\partial n_\mathrm{HOAO}}.
\end{align}
By enforcing $U^\mathrm{RSH-DFTB} \overset{!}{=} U^\mathrm{RSH-DFT}$, we finally end up with the Hubbard $U$ of RSH-DFTB
\begin{widetext}
\begin{align}\label{eq:U-cam-dftb}
U^\mathrm{RSH-DFTB}_A &= \frac{5}{16}\tau_A\left[1 - \frac{1}{2(2l + 1)}\left\{\alpha + \beta \left(1 - \frac{\tau_A^8 + 3\tau_A^6\omega^2 - \tau_A^4\omega^4 + \frac{1}{5}\omega^6\tau_A^2 - \frac{16}{5}\tau_A^7\omega}{(\tau_A^2 - \omega^2)^4}\right)\right\}\right],
\end{align}
\end{widetext}
an equation that should be solved numerically to obtain the decay constant $\tau_A$.

\section{$\mathrm{\Gamma}$-point approximation}\label{sec:gamma-point-approx}
For periodic RSH-DFTB calculations that are restricted to only the $\Gamma$-point, we resort to the TCI scheme exclusively.
In the $\Gamma$-point-only approximation, any phase factors vanish and the $\bm{g}$-summation of Eq.\,\eqref{eq:dHmn-rspace-dftbplus} is carried out in advance, yielding new $\gamma$-integrals
\begin{align}
\tilde{\gamma}_{MN}^\mathrm{TC} &= \sum_{\bm{g}} \gamma_{MN}^\mathrm{TC}(\bm{g} + \bm{l}).
\end{align}
With $\gamma_{MN}^\mathrm{TC}$ truncated spherically and the $\bm{g}$-summation covering the entire crystal, the value of $\tilde{\gamma}_{MN}^\mathrm{TC}$ eventually becomes independent of the shift, $\bm{l}$, which greatly facilitates the pre-tabulation of $\tilde{\gamma}_{MN}^\mathrm{TC}$ for all element combinations.

From this we infer that the exchange Hamiltonian at the $\Gamma$-point $\Delta {\bf H}^{x,\mathrm{lr}}(\bm{\Gamma})$ can be constructed from matrix-matrix multiplications of dense overlap and density matrices
\begin{align}\label{eq:dHmn-rspace-gamma-tc-mm}
\Delta {\bf H}^{x,\mathrm{lr}}(\bm{\Gamma}) &= -\frac{1}{8} \Big\{\big[{\bf S}(\bm{\Gamma}) \Delta {\bf P}(\bm{\Gamma}) {\bf S}(\bm{\Gamma})\big] \odot \tilde{\gamma}^\mathrm{TC} \notag
\\
&+ \big[\big({\bf S}(\bm{\Gamma}) \Delta {\bf P}(\bm{\Gamma})\big) \odot \tilde{\gamma}^\mathrm{TC} \big] {\bf S}(\bm{\Gamma}) \notag
\\
&+ {\bf S}(\bm{\Gamma}) \big[\big(\Delta {\bf P}(\bm{\Gamma}) {\bf S}(\bm{\Gamma}) \big) \odot \tilde{\gamma}^\mathrm{TC} \big] \notag
\\
&+ {\bf S}(\bm{\Gamma}) \big(\Delta {\bf P}(\bm{\Gamma}) \odot \tilde{\gamma}^\mathrm{TC} \big) {\bf S}(\bm{\Gamma}) \Big\},
\end{align}
with $\tilde{\gamma}^\mathrm{TC}$ denoting a super-matrix with the same shape as $S(\bm{\Gamma})$ and $\Delta P(\bm{\Gamma})$.
Element-wise multiplication is indicated by $\odot$, i.e.\ the Hadamard product.
Eq.\,\eqref{eq:dHmn-rspace-gamma-tc-mm} is a direct generalization of the non-periodic algorithm that is already implemented in the DFTB+~\cite{dftb+} code.
Establishing $\tilde{\gamma}^\mathrm{TC}$ is straightforward as all elements of the diatomic block between atoms $M$ and $N$, containing orbitals \{$\mu[M], \nu[N]$\}, takes the same value: $\tilde{\gamma}_{MN}^\mathrm{TC}$, i.e.\ $\tilde{\gamma}^\mathrm{TC}_{\mu[M]\nu[N]} = \tilde{\gamma}_{MN}^\mathrm{TC}$.
This algorithmic solution is appealing, since it is exact in the sense that no integral screening, as described in Section~\ref{sec:screening}, is required and since dense matrix-matrix multiplications can be performed efficiently and are easy to parallelize.
A neighbor-list based algorithm for $\Gamma$-point calculations that includes integral pre-screening is implemented as well. However, for the systems so far tested, we observed a significantly higher efficiency for the matrix-multiplication based algorithm, therefore we refrain from discussing the list-based approach in detail.

Analogously to Eq.\,\eqref{eq:dHmn-rspace-gamma-tc-mm}, the total energy and its gradients, i.e.\ atomic forces, can be expressed in a similar fashion
\begin{align}\label{eq:Gamma-point_forces}
E^{x,\mathrm{lr}}_\Gamma &= -\frac{1}{16} \Big\{\big[{\bf S}(\bm{\Gamma}) \Delta {\bf P}(\bm{\Gamma}) {\bf S}(\bm{\Gamma})\big] \odot \tilde{\gamma}^\mathrm{TC} \notag
\\
&+ \big[\big({\bf S}(\bm{\Gamma}) \Delta {\bf P}(\bm{\Gamma})\big) \odot \tilde{\gamma}^\mathrm{TC} \big] {\bf S}(\bm{\Gamma}) \notag
\\
&+ {\bf S}(\bm{\Gamma}) \big[\big(\Delta {\bf P}(\bm{\Gamma}) {\bf S}(\bm{\Gamma}) \big) \odot \tilde{\gamma}^\mathrm{TC} \big] \notag
\\
&+ {\bf S}(\bm{\Gamma}) \big(\Delta {\bf P}(\bm{\Gamma}) \odot \tilde{\gamma}^\mathrm{TC} \big) {\bf S}(\bm{\Gamma}) \Big\} \Delta {\bf P}(\bm{\Gamma})
\\
F^\alpha_M &= -\frac{\partial E^{x,\mathrm{lr}}_\Gamma}{\partial R^\alpha_M}
\\
&= \frac{1}{2} \sum_{\mu[M]} \sum_{N \neq M} \sum_{\nu[N]} (\partial_{R^\alpha_M} S_{\mu\nu}) \notag
\\
&\times \left[ \Delta {\bf P} {\bf S} \left( \Delta {\bf P} \odot \tilde{\gamma}^\mathrm{TC} \right) + \left( \left( \Delta {\bf P} {\bf S} \right) \odot \tilde{\gamma}^\mathrm{TC} \right) \Delta{\bf P} \right]^\text{sym}_{\nu\mu} \notag
\\
&+ \frac{1}{4} \sum_{\mu[M]} \sum_{N \neq M} \sum_{\nu[N]} (\partial_{R^\alpha_M} \tilde{\gamma}^\mathrm{TC}_{\mu\nu}) \notag
\\
&\times \left[ \left( \Delta {\bf P} {\bf S} \right)^\top \odot \left( \Delta {\bf P} {\bf S} \right) + \left( {\bf S} \Delta {\bf P} {\bf S} \right) \odot \Delta {\bf P} \right]^\text{sym}_{\nu\mu},
\end{align}
with Cartesian coordinates $R^\alpha_M$ of atom $M$, $\alpha \in \{x,y,z\}$ and ${\bf A}^\text{sym} = \frac{1}{2} ({\bf A} + {\bf A}^\top)$ denoting the symmetric part of matrix ${\bf A}$.

\bibliography{references}

\end{document}